%% file: main.tex
\documentclass[11pt,a4paper]{article}

\input{mypackages}

\usepackage[UKenglish]{babel}
\usepackage{float}
\usepackage{microtype}
\usepackage{hyperref}

\theoremstyle{plain}
\newtheorem{theorem}{Theorem}[section]

\newtheorem{proposition}[theorem]{Proposition}

\theoremstyle{definition}
\newtheorem{definition}[theorem]{Definition}

\title{
An Efficient and Uniform CSP Solution Generator Generator\thanks{
This paper is an extended version of a paper published at CP 2025.
}
}

\author{
Ghiles Ziat\thanks{
EPITA Research Lab (LRE), France.
\href{https://ghilesz.github.io/}.
Email: \texttt{ghiles.ziat@epita.fr}.
ORCID: \href{https://orcid.org/0000-0002-9623-9416}.
}
\and
Martin Pépin\thanks{
Université Caen Normandie, ENSICAEN, CNRS, Normandie Univ,
GREYC UMR 6072, F-14000 Caen, France.
\href{https://pepin231.users.greyc.fr}.
Email: \texttt{martin@martin.martin}.
ORCID: \href{https://orcid.org/0000-0003-1892-3017}.
}
}

\date{}

\begin{document}

\maketitle

\begin{abstract}
Constraint-based random testing is a powerful technique which aims at
generating random test cases to verify functional properties of a
program. Its objective is to determine whether a function satisfies a
given property for every possible input. This approach requires
firstly defining the property to satisfy, then secondly to provide a
``generator of inputs'' able to feed the program with the inputs
generated. Besides, function inputs often need to satisfy certain
constraints to ensure the function operates correctly, which makes the
crafting of such a generator a hard task. In this paper, we are
interested in the problem of manufacturing a uniform and efficient
generator for the solutions of a CSP. In order to do that, we propose
a specialized solving method that produces a well-suited
representation for random sampling. Our solving method employs a
dedicated propagation scheme based on the hypergraph representation of
a CSP, and a custom split heuristic called \emph{birdge-first} that
emphasizes the interests of our propagation scheme. The generators we
build are general enough to handle a wide range of use-cases. They are
moreover uniform by construction, iterative and self-improving. We
present a prototype built upon the \textit{AbSolute} constraint
solving library and demonstrate its performances on several realistic
examples.
\end{abstract}

\section{Introduction}
\input{intro}

\section{What is a (Good) Generator?}
\label{sec:generators}
\input{generators}

\section{Constraint-Based Generators}
\label{sec:constraint-generators}
\input{constraint-generators}

\section{Graph Representation for Random Sampling}
\label{sec:graph}
\input{graphs}

\section{Interdimensional Travel}
\label{sec:incr}
\input{incremental}

\section{Constraint Based Testing with GeGen}
\label{sec:gegen}
\input{gegen}

\section{Experiments}
\label{sec:exp}
\input{experiments}

\section{Related Works}
\label{sec:related}
\input{related}

\section{Conclusion}
\label{sec:con}
\input{conclusion}

\bibliographystyle{plain}
\bibliography{biblio}

\end{document}

%% file: mypackages.tex
\usepackage{algorithm}            
\usepackage[noend]{algpseudocode} 
\usepackage{graphicx}
\usepackage{listings}
\usepackage{xcolor}               
\usepackage{url}              
\usepackage{tikz}
\usepackage{todonotes}
\usetikzlibrary{arrows.meta,patterns}
\usepackage{xspace}               
\usepackage{amsmath}              
\usepackage{amssymb}              
\usepackage{amsthm}              
\usepackage{mwe} 
\usepackage{subcaption}

\definecolor{codegreen}{rgb}{0,0.6,0}
\definecolor{codegray}{rgb}{0.5,0.5,0.5}
\definecolor{codepurple}{rgb}{0.58,0,0.82}
\definecolor{backcolour}{rgb}{0.95,0.95,0.92}
\definecolor{darkblue}{rgb}{0.2,0.2,0.4}
\definecolor{darkgreen}{rgb}{0.3,0.5,0.3}

\makeatletter
\DeclareRobustCommand\onedot{\futurelet\@let@token\@onedot}
\def\@onedot{\ifx\@let@token.\else.\null\fi\xspace}
\makeatother

\def\ie{\emph{i.e}\onedot}

\lstdefinestyle{ocaml}{
    backgroundcolor=\color{backcolour},   
    commentstyle=\color{codegreen},
    keywordstyle=\bfseries\color{red},
    numberstyle=\tiny\color{codegray},
    stringstyle=\color{codepurple},
    breakatwhitespace=false,         
    breaklines=true,                 
    captionpos=b,                    
    keepspaces=true,                 
    numbers=left,                    
    numbersep=5pt,                  
    showspaces=false,                
    showstringspaces=false,
    showtabs=false,                  
    tabsize=2,
    identifierstyle=\idstyle,
    basicstyle=\footnotesize,
    morekeywords={type, let, fun, if, then, else, match, with, function},
    moredelim = [s][\color{darkgreen}]{[@@}{]},
    moredelim = [s][\color{gray}]{(\*}{\*)}
}

\makeatletter
\newcommand*\idstyle{%
  \expandafter\id@style\the\lst@token\relax
}
\def\id@style#1#2\relax{%
  \ifcat#1\relax\else
  \ifnum`#1=\uccode`#1%
  \bfseries\color{darkblue}
  \fi
  \fi
}
\makeatother

\newcommand{\GComment}[1]{\textcolor{gray}{\Comment{{\footnotesize #1}}}}
\newcommand\Inner{I}
\newcommand\Outer{O}

\graphicspath{{./imgs/}}

%% file: intro.tex
The objective of this work is to propose a technique for building an
efficient and uniform sampler of solutions from a Constraint
Satisfaction Problem (CSP).

Generating uniformly distributed solutions to a Constraint
Satisfaction Problem can be useful in various applications where
fairness, diversity, unbiased testing, or comprehensive exploration of
solution spaces is required. For resource allocation and fair
division, applications like cloud computing and fair task distribution
rely on uniform sampling to guarantee equity~\cite{Ge&al07}. In
machine learning, uniform solution generation helps creating diverse
training datasets for constraint-based domains~\cite{Ch&al13}.
Cryptography benefits from uniform sampling for unpredictable key
generation~\cite{De&al08}. Also, applications like network testing
leverage uniform sampling for unbiased evaluation of routing
algorithms~\cite{Go&al04}.

Also, crafting a uniform sampler is particularly useful in the
context of Property Based Testing~\cite{FiBi97} (PBT), and more
specifically in random testing in which they are refered to
as \textit{generators}. Random testing is a black-box testing
technique where programs are tested by generating random, independent
inputs, provided by a generator. Results of the output are compared
against software specifications to verify that the test passes or
fails. While generarating inputs uniformly at random can be
straightforward, it is often necessary to generate inputs that satisfy
certain number of prerequisites, while assuring a good coverage of the
input space and maintaining reasonable execution time.
Frameworks \textit{à la} QuickCheck~\cite{Cl&al00} usually deal with this problem by
providing \textit{rejection sampler} combinators: a candidate input is
generated using a base generator and kept if it satisfies some
constraints, or discarded otherwise.

For instance, consider the problem of generating two integers \(x\)
and \(y\), both uniformly drawn from the interval \([0,1000]\), such
that their sum is exactly \(1000\). Using a rejection-sampling
approach, this can be expressed as follows:

\begin{lstlisting}[caption={OCaml code using the QCheck library for generating integer pairs satisfying a constraint using rejection sampling}, label={lst:rejection-example}]
find_example
  (pair (int_bound 1000) (int_bound 1000))
  (fun (x, y) -> x + y = 1000)
\end{lstlisting}

The generator first samples a pair of integers, bouth between $0$ and
$1000$, and then checks whether the sum constraint is satisfied. If it
is not, the candidate is discarded and another pair is
generated. Although this example is relatively simple, only 1001 of
the 1,002,001 possible pairs satisfy the constraint. Consequently, the
probability of accepting a randomly generated candidate is
approximately \(0.1\%\), meaning that, on average, around 1000
candidates must be generated before obtaining a valid one. This
illustrates a fundamental limitation of rejection sampling: as the
proportion of valid solutions decreases, the number of discarded
candidates increases, potentially making generation prohibitively
expensive.  The rejection process is generally repeated a fixed number
of time until a candidate is found or giving up. This approach is
simple, both to use and to implement. However, it might not be
efficient in scenarios where the solution space is sparse or where
constraints are complex.

Despite its simplicity, rejection sampling has an important advantage:
provided that the base generator is uniform, the resulting
distribution over valid solutions is also uniform. Moreover, when the
solution space is sufficiently dense, valid candidates are found
frequently, making rejection sampling an effective and inexpensive
approach.

Constraint solving offers an alternative by providing mechanisms to
explore the solution space while avoiding the repeated generation of
invalid candidates. This can substantially improve generation
efficiency when solutions are sparse or constraints are particularly
restrictive. Nevertheless, this improvement comes at a cost: calling a
constraint solver can introduce an important computational
overhead. When solutions are abundant, this overhead may outweigh the
benefits of guided generation. Despite this cost, several works have
demonstrated that constraint-solving techniques can make the overall
process more efficient in practice~\cite{Zi&al21,Cl&al14}.  An
additional difficulty is that efficiently finding solutions does not
necessarily imply sampling them uniformly. Many existing
approaches~\cite{Ch&al13,Pe&al22,Va&al21} settle for quasi-uniform
distributions or weaker notions of uniformity, trading sampling
guarantees for efficiency. Achieving strict uniformity while retaining
reasonable generation performance therefore remains a challenging
problem.

QuickCheck's rejection sampling provides a straightforward way to
create a generator, albeit one that may be slow. In contrast,
constraint-solving techniques adopt a more computationally intensive
process to construct a fast generator.  Our proposed incremental
approach seeks to strike a balance between these, aiming to achieve an
optimal trade-off between the speed of generator construction and
runtime efficiency.  This balance is particularly beneficial in
scenarios where there is little or no prior information about how many
times the generator will be used. In such cases, the upfront cost of
constructing a highly optimized generator through constraint solving
may be unjustified if the generator is used only a handful of
times. Conversely, relying solely on a quickly constructed, but slow
generator, may result in suboptimal performance if the generator is
reused extensively.  By incrementally refining the generator as
needed, our approach adapts to different usage patterns, providing a
flexible and efficient solution regardless of the frequency of
generator usage. This scalability is achieved through our novel
selection structure, which is designed to handle more complex forms,
such as non-fixed-size structures (e.g., list matrices), a challenge
that many existing approaches fail to address.  Instead of immediately
generating examples, our method abstracts away the generation process,
allowing for dynamic adjustment and refinement of the generator based
on constraints. It thus incrementally
\textit{generates} a generator that encapsulates the desired properties and
constraints, hence the title of this paper.

\subsection{Contributions}

\begin{itemize}
\item A general method to produce fast and self-improving uniform
  samplers of CSP solutions using a dedicated solving algorithm for
  random generation.
\item A graph based split heuristic and propagation scheme, well fit
  to partition a problem into independent sub-problems.
\item A heuristic Huffman-like representation for the solutions which
  minimizes the cost of uniform choices.
\item A property based testing API, \textit{à la quickcheck}, allowing for a
  transparent usage of our hybrid approach between rejection sampling and
  constraint based solving. The implementation is available at:\\
  {\small\url{https://osf.io/u4r5q/files/osfstorage?view_only=84af8ee65a6c495d98cb7b7bfad8c54b}}
\end{itemize}

\subsection{Outline}
This paper is organized as follows: Section~\ref{sec:generators}
defines the concept of generators in general and gives some insight of
what is expected from a \textit{good}
generator. Section~\ref{sec:constraint-generators} introduces the main
mechanisms needed to build a constraint-based
generators. Section~\ref{sec:graph} is the main contribution of this
work: it addresses the problem of building
a \textit{propagation-exploration} scheme well-suited for the design
of a constraint based generator. Section~\ref{sec:exp} presents our
implementation and show its performances on some benchmark. Finally,
Section~\ref{sec:related} presents the related works and
Section~\ref{sec:con} summarizes our work and discusses its future
continuations.

%% file: generators.tex
CSP's search spaces are defined by variables to which are associated
finite bound domains. A point in this search space in generally called
an instance, but for our purpose we will call those samples.

\begin{definition}[Sample]
Given a set of variables $\mathcal{V}$, and a set of domains
$\mathcal{D}$ denoting the possible values of variables, a sample is
a total mapping from variables to values. We note the set of samples
$\mathcal{S}=\mathcal{V} \to \mathcal{D}$
\end{definition}

We distinguish two kinds of samples, the ones that satisfies all the
constraint of a given CSP \ie the solutions, and the ones that
violate at least one constraint.

\begin{definition}[Solution Generator]
Given a CSP $(\mathcal{V}, \mathcal{D}, \mathcal{C})$, a solution
generator is a function $g$ that takes a random state and produces a
solution $s \in S$, such that:
$$\forall r, \forall c \in \mathcal{C}, \quad c(g(r)) \text{ holds}$$
where the first quantification is over all the possible random states~$r$.
\end{definition}

This definition ensures that every generated sample satisfies all
constraints of the CSP. Moreover, the use of random states\footnote{We
do not specify the actual representation of random states, as it is
irrelevant to our discussion and depends on the underlying
implementation of the generator.} is needed so that the generation
process can be made reproducible by controlling or restoring the
state. Good generators should be:

\begin{itemize}
\item \textbf{Correct:} They should respect the constraints they
  are subject to. For a data type representing positive integers, a
  generator should ensure that it only produces positive integers.

\item \textbf{Uniform:} They should be able to thoroughly explore the input
  space. In other lines of work, diverseness is ensured by building surjective
  generators (\ie every possible solution can be generated with
  \textit{a non-zero probability}~\cite{Dy&al05}). Here we have a stronger
  requirement: every possible solution must be generated with \textit{the same
  probability}\footnote{Some
  framework focus on building corner-case generators, generally
  defined in an \textit{ad-hoc} fashion, in which case uniformity is
  irrelevant. This is a complementary approach that we are not focused on in this work.}.

\item \textbf{Efficient:} Testing time includes generating time,
  since test cases are generated dynamically. Hence, to be of
  practical use, generators must maintain reasonable performance, in
  particular in large codebases that run tests frequently.
\end{itemize}

Property-based testing frameworks \textit{à la} QuickCheck generally resort to
rejection sampling to produce random values that meet specific
constraints. While this technique meets the first two of the above
requirements, it can lead to a significant overhead, in particular
when a large portion of generated values are rejected. The impact on
performances depends on the efficiency of the rejection process and
the likelihood of rejection. We briefly recall how it works.

\subsection{Rejection Sampling}

In the most general sense, rejection sampling is the approach that consists in
generating samples in a super-set of the set of objects we are interested in,
repeatedly until a sample lying in the set of interest is found.
For instance, in order to generate a point inside the disk of radius~$1$
in~${\mathbb R}^2$, one could generate points~$(X,Y)$ in the square~$[-1,1]^2$
until~$X^2 + Y^2 \leq 1$.
In a constraint solving context, this means generating samples within the bounds
of the problem until a sample that satisfy the constraints is
found.
Algorithm~\ref{genNtest} illustrates this method.

\begin{algorithm*}[ht]
  \begin{algorithmic}[1]
    \Function{sample}{$\mathcal{V}$,$\mathcal{D}$,$\mathcal{C}$}
    \State candidate $\leftarrow$ \texttt{spawn}($\mathcal{V}$,$\mathcal{D}$)\label{line:spawn}
      \Comment{draw at random in~$(\mathcal V, \mathcal D)$}
      \If{\texttt{sat}(candidate, $\mathcal{C}$)}\label{line:sat}
        \State \Return candidate
      \Else{
        \State \Return{sample($\mathcal{V}$,$\mathcal{D}$,$\mathcal{C}$)}
      }
      \EndIf
      \EndFunction
  \end{algorithmic}
  \caption{Rejection sampling procedure\label{genNtest}}
\end{algorithm*}

This algorithm takes as input a set of variables $\mathcal{V}$, their associated
range of values $\mathcal{D}$, and a set of constraints $\mathcal{C}$.
The \texttt{spawn} function generates a candidate at random (for some
probability distribution~$\mathbb P$) from the given variables and their
domains.
Then, the \texttt{sat} function evaluates whether the generated candidate
satisfies all the constraints in $\mathcal{C}$.
The procedure is called recursively until a solution is found.

\begin{proposition}[Termination of Alg.~\ref{genNtest}]
  \label{prop:rej:termination}
  Let~$A$ denote the set of solutions of~$\mathcal C$. If the
  probability~$\mathbb P(A)$ that a random sample given by
  ${\normalfont\texttt{sample}}(\mathcal V, \mathcal D)$ lies in $A$ is non-zero, then
  Algorithm~\ref{genNtest} terminates with probability 1.
  Furthermore, its number of iterations follows a geometric distribution
  (supported by~$\{1, 2, 3, \ldots\}$) of parameter~$\mathbb P(A)$.
\end{proposition}

\begin{proof}
  Let~$N$ denote the random variable that counts the number of iterations of the
  algorithm. This variable takes its values in~$\{1, 2, 3, \ldots\} \cup
  \{\infty\}$.
  Because of the recursive nature of Algorithm~\ref{genNtest} and because every
  call to \texttt{sample} is independent, we have the following recursive
  equation on the distribution of~$N$:
  \begin{equation*}
    \forall n > 0, \mathbb P(N = n) = \begin{cases}
      \mathbb P(A)                        &\text{if $n=1$} \\
      (1 - \mathbb P(A)) \mathbb P(N=n-1) & \textrm{otherwise}.
    \end{cases}
  \end{equation*}
  It follows that~$\mathbb P(N = n) = (1-\mathbb P(A))^{n-1} \mathbb P(A)$ which
  is the claimed distribution. As a consequence, we have that~$\mathbb P(N =
  \infty) = 0$.
\end{proof}

Characterising the law of the number of iterations allows to predict its typical
value, which is essential for average-case complexity.
The expected number of iterations of Algorithm~\ref{genNtest}
is~$\frac{1}{p_{\textrm{sat}}}$.
Another key observation is that Algorithm~\ref{genNtest} implements the
probability distribution~$\mathbb P$ \emph{conditioned} to only draw elements of
the solution space~$A$, which is formalised in Proposition~\ref{genNtest:dist}
below.

\begin{proposition}[Distribution of the output Alg.~\ref{genNtest}]
  \label{genNtest:dist}
  Let~$\mathbb P$ denote the probability distribution implemented by the
  {\normalfont\texttt{spawn}} function, let~$A$ denote the solution set of the problem,
  let~$X$ be the (random) output of Algorithm~\ref{genNtest} and let~$B
  \subseteq A$ be a measurable subset of~$A$.
  The we have that the probability that Algorithm~\ref{genNtest} yields an
  element of~$B$ is
  \begin{equation*}
    \mathbb P_{\textsc{sample}}(X \in B) = \frac{\mathbb P (B)}{\mathbb
    P(A)} = \mathbb P(A|B).
  \end{equation*}
\end{proposition}

\begin{proof}
  Note that Algorithm~\ref{genNtest} only produces elements of~$A$ by
  construction.
  By induction, it returns an element of~$B$ if either
  \begin{itemize}
    \item \texttt{sample}$(\mathcal V, \mathcal D)$ returns an element of~$B$
      at line~\ref{line:spawn}
    \item or \texttt{sample}$(\mathcal V, \mathcal D)$ returns a candidate that
      does not belong to~$A$ and the recursive call to \textsc{sample} produces
      an element of~$B$.
  \end{itemize}
  It follows that
  \begin{equation*}
    \mathbb P_{\textsc{sample}}(X \in B) =
    \mathbb P(B) + (1 - \mathbb P(A)) \mathbb P_{\textsc{sample}}(X \in B)
  \end{equation*}
  which yields the desired result.
  Note that~$\mathbb P(B) = \mathbb P(B \cap A)$ since~$B \subseteq A$.
\end{proof}

A straightforward consequence of Proposition~\ref{genNtest:dist}, is that if the
\texttt{spawn} function draws uniform samples in the domain~$\mathcal D$, then
Algorithm~\ref{genNtest} is a uniform sampler of solutions.
The contribution of the present paper is precisely to devise a good
\texttt{spawn} function that can guarantee that this algorithm is uniform while
providing good performance, that is a high enough value for~$\mathbb P(A)$.

%% file: constraint-generators.tex
In this article, we use a constraint-based approach for random
sampling. To achieve this, we rely on the general abstract solving
method described in \cite{Pe&al13}, which we summarize here.

\begin{algorithm}[ht]
  \begin{algorithmic}[1]
    \Function{solve}{$\mathcal{X},\mathcal{D},\mathcal{C}$}
    \GComment{$\mathcal{X}$:variables, $\mathcal{D}$: domains, $\mathcal{C}$: constraints}
    \State $I \leftarrow \emptyset$ \GComment{inner solutions}
    \State $\Outer \leftarrow \{$\texttt{init}($\mathcal{D})\}$  \GComment{outer solutions}
    \While{$\Outer \neq \emptyset$}
      \State $e \leftarrow$ \texttt{select}($\Outer$)
      \State $e' \leftarrow$ \texttt{filter}$(e,\mathcal{C})$
      \If{$e' \neq \bot$}
        \If{\texttt{solution}$(e',\mathcal{C})$}
          \State $\Inner \leftarrow \Inner \cup \{e'\}$
          \Else{\State $\Outer \leftarrow \Outer \cup \texttt{split}(e)$}
        \EndIf
        \EndIf
        \EndWhile \State \Return{$\Inner,\Outer$} \EndFunction \end{algorithmic} \caption[Abstract
    solving]{Abstract solving method\label{abssolvegen}}
\end{algorithm}

Algorithm~\ref{abssolvegen} constructs a cover of the solution space
using abstract elements (e.g., boxes, octagons, polyhedra, etc.). This
cover consists of two sets: inner elements ($\Inner$) and outer elements
($\Outer$). The set $\Inner$ under-approximates the solution set, while $\Inner \cup
\Outer$ over-approximates it. The algorithm starts by initializing an
abstract element and inserting it into $\Outer$. It then iterates through
the following steps: an element from $\Outer$ is selected, filtered, and,
if it satisfies the constraints, added to $\Inner$. If it does not, it is
split into sub-elements that are reinserted into $\Outer$.

As presented, this algorithm may not terminate. In practice, various
termination criteria can be employed, such as limiting the size of the
elements considered or the depth of the solving tree.  Note that, in
general, split elements can overlap without preventing the algorithm
from terminating and producing correct results. However, in our case,
ensuring they are non-overlapping is necessary for uniformity, as we
will discuss later.

Although the algorithm is parametric with respect to the
representation being used, we only use boxes in our implementation.
Recall that given variables $v_1, \dots, v_n$ over finite continuous
domains $d_1, \dots, d_n$, a box is defined as a Cartesian product of
intervals within $d_1 \times \dots \times d_n$. Boxes offer several convenient
properties for our purposes. First, they can be split into
non-overlapping sub-boxes, allowing the search space to be partitioned
without introducing overlaps. Second, their volume can be computed
efficiently as the product of the lengths of their constituent
intervals. Finally, they enable straightforward and efficient uniform
random sampling: since a box is a Cartesian product of intervals, a
uniform sample can be obtained by independently drawing a uniform
random value from each interval.

{\centering
\begin{figure}[ht]
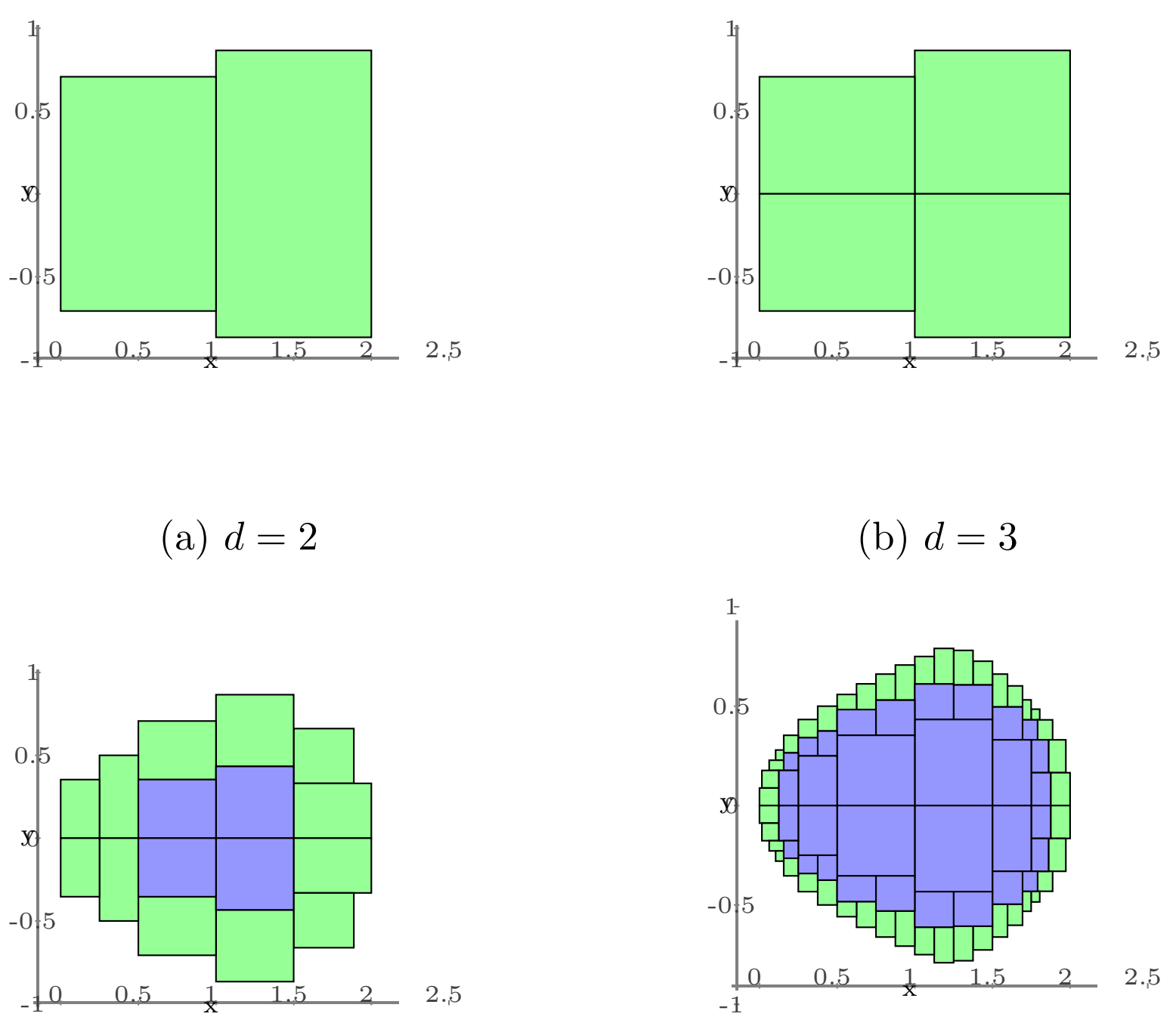

\begin{tabular}{cc}
  \begin{subfigure}{0.45\textwidth}\centering
    \include{imgs/mickey_box_2.tex}
    \caption{$d=2$}\label{fig:taba}
  \end{subfigure}&
  \begin{subfigure}{0.45\textwidth}\centering
    \include{imgs/mickey_box_3.tex}
    \caption{$d=3$}\label{fig:tabb}
    \end{subfigure}\\ \newline
  \begin{subfigure}{0.45\textwidth}\centering
    \include{imgs/mickey_box_5.tex}
    \caption{$d=5$}\label{fig:tabc}
  \end{subfigure}&
  \begin{subfigure}{0.45\textwidth}\centering
    \include{imgs/mickey_box_8.tex}
    \caption{$d=8$}\label{fig:taba2}
  \end{subfigure}\\
\end{tabular}
\caption{Solving state obtained for depth $d$ of resolution for a problem with two variables constrained by $x^2+4y^2-4 \leq 0$ and $2y^2-x \leq 0$. Darker boxes indicate inner elements.}
\label{tab:mytable}
\end{figure}
}

\subsection{Constrained Based Sampling}
In \cite{Zi&al21}, authors build upon this constraint-solving method
an algorithm for uniform sampling under constraints, as shown in
Algorithm~\ref{sample}.  It repeatedly selects an element $e$ from
either the inner or outer sets using a \texttt{select}
function. Here, \texttt{select} performs a weighted choice based on
the volume of the elements, with the largest elements having the best
chance of being chosen. We discuss in the next section the design of a
data structure that enables an efficient implementation of this
function. The algorithm generates a candidate value $i$ within $e$. If
$e$ belongs to the inner set or $i$ satisfies the constraints
$\mathcal{C}$, the algorithm returns $i$; it discards the sample and
repeats these steps, including the weighted choice, otherwise.

\begin{algorithm}[H]{
    \begin{algorithmic}[1]
      \Function{generate}{$\Inner, \Outer, \mathcal{C}$}
        \While{true}
          \State $e \gets \texttt{select}(\Inner,\Outer)$\label{line:selectbox}
          \State $i \gets \texttt{spawn}(e)$\label{line:sampleinbox}
          \If{$e \in \Inner \lor \texttt{sat}(i,\mathcal C)$}\label{line:rejection}
            \Return $i$
          \EndIf
        \EndWhile
      \EndFunction
    \end{algorithmic}
  }
  \caption{Random Sampling for Covers\label{sample}}
\end{algorithm}

This algorithm ensures uniform sampling under three conditions: the elements in
$\Inner \cup \Outer$ must not overlap, points within each element $e$ must be
sampled uniformly at random, and the probability of selecting an element $e$
must be proportional to its volume ($P(e) =\frac{v(e)}{\sum_{e' \in \Inner \cup
\Outer} v(e')}$).

Note that the two notions of \emph{uniformity} and \emph{volume} have a distinct
meaning in a discrete and in a continuous context.
In the discrete setup where the total number of points in the domain~$\mathcal D$
is finite (though usually very large), the volume of~$e$ refers to its
cardinality and sampling uniformly at random within~$e$ thus means that every
point~$i \in e$ has a probability~$\frac{1}{v(e)}$ to be drawn.
In the continuous setup, the elements~$e$ are product of intervals and their
volume is the usual notion of volume in~$\mathbb R^n$ (their Lebesgue measure).
Sampling uniformly at random within $e \subseteq \mathbb R^n$ then means
sampling according to~$\frac{dx_1 dx_2 \cdots dx_n}{v(e)}$.

These conditions ensure that the sampling process remains uniform
across the domain, giving all points in the union of elements an equal
likelihood of being chosen.
\begin{proposition}\label{uniformityclaim}
  Assuming that the set of solutions satisfying the constraints~$\mathcal C$ has
  positive measure (in particular, it is non-empty), Algorithm~\ref{sample}
  samples solutions uniformly at random.
\end{proposition}
While uniformity was an implicit goal in \cite{Zi&al21}, we provide a formal
proof to establish it rigorously.

\begin{proof}
  Let \(\mathcal{S}\) be the set of all solutions satisfying constraints
  \(\mathcal{C}\).
  By construction we have that~$\mathcal S$ is contained in the union of all
  elements of~$\Inner \cup \Outer$, that is
  \begin{equation*}
    \mathcal S \subseteq \bigcup_{e \in \Inner \cup \Outer} e.
  \end{equation*}
  Note that lines~\ref{line:selectbox} and~\ref{line:sampleinbox} in
  Algorithm~\ref{sample} generate a random element of~$\Inner \cup \Outer$ and
  note that the logic implemented by the loop and by the check at
  line~\ref{line:rejection} is that of a rejection sampler.
  Thus, it follows that if the random sample~\texttt{i} is uniform
  within~$\Inner \cup \Outer$ then Algorithm~\ref{sample} is a uniform sampler
  for~$\mathcal S$ by Proposition~\ref{genNtest:dist}.

  Assume that for any element~$e$ of~$\Inner \cup \Outer$, the function call
  \texttt{spawn($e$)} generates a uniform element of $e$, \ie{}, for any
  measurable subset of~$B \subseteq e$ of~$e$, we have $\mathbb
  P({\normalfont\texttt{spawn(e)}} \in B) = \frac{v(B)}{v(e)}$ where~$v(B)$
  and~$v(e)$ are the volumes of~$B$ and~$e$.

  Then, for any measurable subset~$A \subseteq \mathcal S$ of~$\mathcal S$, we
  have that
  \begin{align*}
    \mathbb P(\texttt{spawn}(\texttt{select}(\Inner, \Outer)) \in A)
    &= \sum_{e \in \Inner \cup \Outer}
        \mathbb P(\texttt{select}(\Inner, \Outer) = e) \mathbb
        P(\texttt{spawn}(e) \in (A \cap e)) \\
    &= \sum_{e \in \Inner \cup \Outer}
         \frac{v(e)}{\sum_{e' \in \Inner \cup \Outer} v(e')} \frac{v(A\cap
         e)}{v(e)} \\
    &= \frac{v(A)}{\sum_{e' \in \Inner \cup \Outer} v(e')}\cdot
  \end{align*}
  Note that the last line in the above equation holds because none of the
  elements of~$\Inner \cup \Outer$ intersect.
  This proves that the element~$i$ at line~\ref{line:sampleinbox} in the
  Algorithm is uniform and thus the whole algorithm is a uniform sampler
  for~$\mathcal S$.
\end{proof}

Recall that by Proposition~\ref{prop:rej:termination}, the number of
iterations of Algorithm~\ref{sample} follows a geometric distribution
of parameter~$\frac{v(\mathcal S)}{v(\Inner \cup \Outer)}$.  In
expectation, this amounts for~$\frac{v(\Inner \cup \Outer)}{v(\mathcal
S)}$ iterations per call.  One one hand, refining the solving process
can potentially reduce the rejection rate by reducing the
volume~$\Inner \cup \Outer$ (though not always) and it never increases
it.  However, this on the other hand inevitably increases the
selection time.  At some point, this may even become
counterproductive; when the increase in selection time exceeds the
gains obtained from reducing the rejection rate, further refinement is
no longer beneficial. We confirm this experimentally afterwards.  Our
goal is to find the best trade-off between constraint solving and
rejection sampling, minimizing unnecessary exploration while ensuring
efficiency.

\subsection{Efficient selection of an abstract element}
In order to implement the \texttt{select} function, we need a data
structure for storing a collection of abstract elements that supports:
\begin{itemize}
  \item efficient sampling of an elements with probability proportional to
    its volume;
  \item and (we will see later), an efficient way to replace an abstract element
    with a collection of smaller elements.
\end{itemize}
In \cite{Zi&al21}, sorted list in decreasing order of volume are used
for element selection, so that the elements with highest probability
are met faster during sampling. However, if elements are of the same
size, this approach offers no advantage, as all selections become
equally likely. Also, they do not require to update their structure in
their work.  We can do better by arranging these elements in a binary
tree.  The idea is to store the abstract elements at the leaves of the
tree and to maintain, in every node, the sum of the volumes of all the
leaves below that node.  Using such a data structure, drawing a
abstract element proportional to its volume corresponds to drawing a
uniform real variable between~$0$ and the total volume of the tree,
and to recursively descend in the tree, choosing between the left and
right child based on their weights.  This is illustrated in
Algorithm~\ref{alg:draw:tree}.

\begin{algorithm*}[ht]
  \begin{algorithmic}[1]
    \Function{randomTreeSelect}{$T$}
      \State$r \gets \text{\Call{Unif}{$0$, $\mathrm{volume}(T)$}}$
      \State\Return\Call{treeSelect}{$r$, $T$}
    \EndFunction
  \end{algorithmic}
  \smallskip
  \hrule
  \smallskip
  \begin{algorithmic}[1]
    \Require{$0 \leq r < \mathrm{volume}(T)$}
    \Function{treeSelect}{$r$,$T$}
      \If{$T$ is a leaf}
        \Return the abstract element stored in~$T$
      \EndIf
      \State$(T_L, T_R) \gets \text{the children of~$T$}$
      \If{$r < \mathrm{volume}(T_L)$}
        \Return\Call{treeSelect}{$r$, $T_L$}
      \Else{}
        \Return\Call{treeSelect}{$r - \mathrm{volume}(T_L)$, $T_R$}
      \EndIf
    \EndFunction
  \end{algorithmic}
  \caption{Random generation of an abstract element using the tree data structure\label{alg:draw:tree}}
\end{algorithm*}

Here, the cost of selecting an element is proportional to the length of the path
from the root to the selected leaf.
Given the volume of every abstract element (and thus their probability of being
drawn), there is an optimal way to arrange these elements in the tree in order
to minimise the expected cost of the generation.
Information theoretic results tell us that the expected path length between the
root and the sample is lower bounded by the entropy of the probability
distribution, that is
\begin{equation*}
  \mathcal H = \sum_x p_x \ln\frac{1}{p_x}
\end{equation*}
where the sum ranges over the abstract elements stored in the tree, and~$p_x$ is
the sampling probability of element~$x$.
The optimal way to organise the tree in order to remain close to this lower
bound is to use a \emph{Huffman tree}~\cite{Huffman1952}:
\begin{itemize}
  \item start from a collection of leaves,
  \item iteratively pair the two smallest elements of the collection as a binary
    node,
  \item this process terminates when every leaves belong to the same tree.
\end{itemize}
A tree built this way has the property that its expected path length to a leaf
is at most~$\mathcal H + 1$.

Unfortunately, the good properties of Huffman trees are hard to maintain
efficiently when we update the collection of abstract elements, which we need
for in the adaptive algorithm presented in Section~\ref{sec:incremental}.
To circumvent this issue, we use the following heuristic:
\begin{enumerate}
  \item before doing any sampling, we do a first solving pass until a certain
    depth;
  \item\label{item:huffman} after this pre-processing, we construct a Huffman
    tree~$T$ based on the volumes of the resulting elements;
  \item then, during the iterative sampling process, every time we need to split
    an abstract element~$e$ into a collection of smaller elements~$(e_1, e_2,
    \ldots, e_p)$, we construct a Huffman tree~$T'$ for~$(e_1, e_2, \ldots,
    e_p)$ and we replace the leaf~$e$ in~$T$ with~$T'$.
\end{enumerate}
It is worth noting that the replacement of the last sampled leaf can be
optimised by keeping a pointer to this last leaf, rather than traversing the
tree a second time to find it.

The initial Huffman tree~$T$ constructed at step~\ref{item:huffman} thus
potentially evolves away from its optimal shape as we update leaves.
However, since the elements that have the highest probability to be selected
(and thus split) are the bigger ones, we expected our heuristic to maintain some
balance in the tree.

\subsection{Impact of the Data Structure on Sampling Performance}
To evaluate the impact of the data structure used for element
selection, we measure the sampling throughput obtained after different
amounts of constraint solving. For each solving depth, we first
perform the corresponding number of solving rounds to construct a
cover of the solution space, and then build a generator from this
cover using either a list or a Huffman tree. We then measure the
number of successfully generated samples per CPU second, averaging
over multiple runs to reduce measurement noise. This experiment allows
us to observe how the increasing cost of maintaining and selecting
elements affects the overall sampling performance as the solving
process progresses.

For clarity, we only present the results for one representative
problem, involving two variables and a set of nonlinear
constraints. The same general trends can be observed across the other
problems we experimented with.

\begin{figure}[ht]
\includegraphics[scale=0.4]{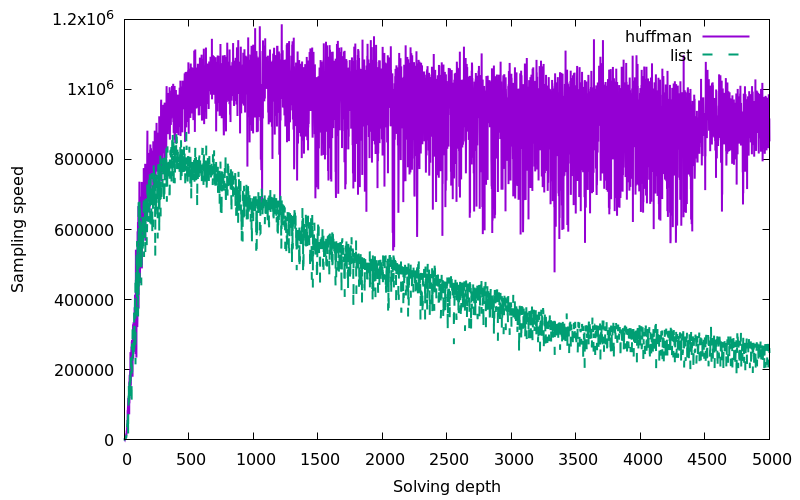}
\caption{Generation speed given samples/second, with both the list and huffman representation}
\label{fig:genspeed}
\end{figure}

Figure~\ref{fig:genspeed} compares the sampling speed of the two data
structures as the solving depth increases. Both approaches initially
benefit from increasing the solving depth, with sampling speed rising
rapidly before reaching a peak. However, their behaviour differs
considerably beyond this point.

For the list-based implementation, performance progressively
deteriorates as the solving depth increases. After reaching
approximately 800,000 samples per second at a depth of around 400, the
sampling speed steadily decreases, eventually falling below 300,000
samples per second at the maximum depth considered. This confirms that
the increasing cost of managing and selecting elements in the list
eventually outweighs the benefits of refining the solution space.

In contrast, the Huffman-based implementation consistently achieves
higher sampling speeds. After its initial increase, its performance
remains relatively stable, fluctuating around one million samples per
second throughout most of the experiment. Although some degradation
may be observed at greater depths, it is considerably less pronounced
than with the list-based implementation. This suggests that the
Huffman tree successfully limits the cost of element selection, even
as the number of elements grows.

Overall, these results demonstrate the advantage of the Huffman-based
representation over the list-based approach, particularly at greater
solving depths. While the list suffers from a substantial decrease in
sampling performance, the Huffman implementation maintains a
relatively high throughput. However, this advantage does not come
without limitations: as the solving depth increases, the growing
number of abstract elements inevitably increases the cost of
selection, potentially making further refinement counterproductive. In
the following section, we introduce a heuristic designed to mitigate
this issue by avoiding unnecessary refinement and, in particular,
oversplitting.

Moreover, our benchmarks, presented in Section~\ref{sec:exp}, seem to
show that this data structure performs well in practice.

\subsection{Incremental and Adaptive Generators}\label{sec:incremental}

From a practical point of view, we cannot really replace an
inefficient sampler with a generator based on an expensive constraint
solving mechanism. This would simply replace a slow generation speed,
by a faster one preceded by a huge overhead due to solving time. This
is especially prejudicial when the generator is only used a handful of
time.

Iterative solvers are commonly used when dealing with complex
constraint satisfaction problems where finding an exact solution is
computationally infeasible or impractical. Instead of attempting to
solve the entire problem in one step, iterative solvers work
incrementally, refining the solution repeatedly until a certain
termination criterion is met.

We therefore propose to use constraint resolution mechanisms
parsimoniously: first, we will target in priority on certain parts of
the problem, identified on the fly, whose filtering can greatly
improve the performance of random generation. Second, we try to
amortize the cost of the resolution steps (filtering and exploration)
during the generation. For example, every $n$ time a generator is
called, we can perform a resolution step. Thus, the more a generator
is used, the more it improves. To achieve this, the generator embeds
an internal solving state that evolves with each step. This state
retains information from previous resolution steps, allowing the
generator to improve progressively rather than starting from scratch
each time.

Algorithm~\ref{abssolvegen} revolves around two primary decisions:
selecting which element to refine and determining how to refine it.
On the one hand, intuitively, the focus should be on larger elements
with higher rejection rates, as they are more critical to the
rejection rate and, thus, the efficiency of the whole process. On the
other hand, sampling also requires choosing an element, with larger
elements being more likely to be selected. This leads to the question
of whether sampling's focus on larger elements can be used to help
guide the solving process.

For each element in our cover, we track the number of times it was
selected and record the number of times it successfully produced a
sample. We also track the total number of successes and failures for
the whole cover. When our algorithm fails to produce a valid solution
within a selected element, we must then decide whether it is necessary
to split it or not. To do this, we base our choice on its acceptance
rate. If it falls below the global acceptance rate, we proceed to
refine that element, meaning we split it and replace it within the
cover with the resulting sub-elements. Otherwise, it is left
unchanged.  This dynamic adjusting of the refinement process can lead
to a faster convergence and avoids over-splitting, which can
deteriorate the element selection procedure. This is illustrated by
Algorithm~\ref{samplesolve}.

\begin{algorithm}[H]{
    \begin{algorithmic}[1]
	    \Function{generate}{inner,outer,$\mathcal{C}$}
      \While{true}
      \State \texttt{e} $\leftarrow $ select(inner,outer)
      \State \texttt{i} $\leftarrow $ spawn(e)
      \If{$e \in $ inner $\lor$ sat($i$,$\mathcal{C}$)}
      \State \Return $i$
      \ElsIf{rate(e) < global(inner,outer)}
      \State refine e
      \EndIf
      \EndWhile
      \EndFunction
    \end{algorithmic}
  }
  \caption{Sampling-Guided solving\label{samplesolve}}
\end{algorithm}

%% file: imgs/mickey_box_2.tex
\begin{tikzpicture}[scale=0.008]
\def\sx{1}
\def\sy{0.8}
\definecolor{R150G255B150}{RGB}{150,255,150};
\fill[R150G255B150] (200.000000*\sx, 35.141017*\sy) -- (200.000000*\sx, 427.739200*\sy) -- (370.000000*\sx, 427.739200*\sy) -- (370.000000*\sx, 35.141017*\sy) -- (200.000000*\sx, 35.141017*\sy) -- cycle;
\definecolor{black}{RGB}{0,0,0};
\draw[black] (200.000000*\sx, 35.141017*\sy) -- (200.000000*\sx, 427.739200*\sy) -- (370.000000*\sx, 427.739200*\sy) -- (370.000000*\sx, 35.141017*\sy) -- (200.000000*\sx, 35.141017*\sy) -- cycle;
\fill[R150G255B150] (30.000000*\sx, 71.162571*\sy) -- (30.000000*\sx, 391.717645*\sy) -- (200.000000*\sx, 391.717645*\sy) -- (200.000000*\sx, 71.162571*\sy) -- (30.000000*\sx, 71.162571*\sy) -- cycle;
\draw[black] (30.000000*\sx, 71.162571*\sy) -- (30.000000*\sx, 391.717645*\sy) -- (200.000000*\sx, 391.717645*\sy) -- (200.000000*\sx, 71.162571*\sy) -- (30.000000*\sx, 71.162571*\sy) -- cycle;
\definecolor{gray230}{RGB}{230,230,230};
\definecolor{gray128}{RGB}{128,128,128};
\fill[gray128] (400.000000*\sx, 8.005554*\sy) -- (400.000000*\sx, 4.541452*\sy) -- (0.000000*\sx, 4.541452*\sy) -- (0.000000*\sx, 8.005554*\sy) -- cycle;
\fill[gray128] (6.500000*\sx, 462.380216*\sy) -- (3.500000*\sx, 462.380216*\sy) -- (3.500000*\sx, 0.500000*\sy) -- (6.500000*\sx, 0.500000*\sy) -- cycle;
\draw[gray128] (30.000000*\sx,2.809402*\sy) -- (30.000000*\sx, 9.737605*\sy);
\definecolor{gray64}{RGB}{64,64,64};
\node[align=center,text=gray64,font=\tiny] at (30.000000*\sx, 17.820509*\sy) { 0 };
\draw[gray128] (115.000000*\sx,2.809402*\sy) -- (115.000000*\sx, 9.737605*\sy);
\node[align=center,text=gray64,font=\tiny] at (115.000000*\sx, 17.820509*\sy) { 0.5 };
\draw[gray128] (200.000000*\sx,2.809402*\sy) -- (200.000000*\sx, 9.737605*\sy);
\node[align=center,text=gray64,font=\tiny] at (200.000000*\sx, 17.820509*\sy) { 1 };
\draw[gray128] (285.000000*\sx,2.809402*\sy) -- (285.000000*\sx, 9.737605*\sy);
\node[align=center,text=gray64,font=\tiny] at (285.000000*\sx, 17.820509*\sy) { 1.5 };
\draw[gray128] (370.000000*\sx,2.809402*\sy) -- (370.000000*\sx, 9.737605*\sy);
\node[align=center,text=gray64,font=\tiny] at (370.000000*\sx, 17.820509*\sy) { 2 };
\draw[gray128] (455.000000*\sx,2.809402*\sy) -- (455.000000*\sx, 9.737605*\sy);
\node[align=center,text=gray64,font=\tiny] at (455.000000*\sx, 17.820509*\sy) { 2.5 };
\draw[gray128] (2.000000*\sx,4.773442*\sy) -- (8.000000*\sx, 4.773442*\sy);
\node[align=center,text=gray64,font=\tiny] at (5.000000*\sx, 4.773442*\sy) { -1 };
\draw[gray128] (2.000000*\sx,118.106775*\sy) -- (8.000000*\sx, 118.106775*\sy);
\node[align=center,text=gray64,font=\tiny] at (5.000000*\sx, 118.106775*\sy) { -0.5 };
\draw[gray128] (2.000000*\sx,231.440108*\sy) -- (8.000000*\sx, 231.440108*\sy);
\node[align=center,text=gray64,font=\tiny] at (5.000000*\sx, 231.440108*\sy) { 0 };
\draw[gray128] (2.000000*\sx,344.773442*\sy) -- (8.000000*\sx, 344.773442*\sy);
\node[align=center,text=gray64,font=\tiny] at (5.000000*\sx, 344.773442*\sy) { 0.5 };
\draw[gray128] (2.000000*\sx,458.106775*\sy) -- (8.000000*\sx, 458.106775*\sy);
\node[align=center,text=gray64,font=\tiny] at (5.000000*\sx, 458.106775*\sy) { 1 };
\node[align=center,text=black,font=\tiny] at (200.000000*\sx, 0.500000*\sy) { x };
\node[align=center,text=black,font=\tiny] at (0.000000*\sx, 231.440108*\sy) { y };
\end{tikzpicture}

%% file: imgs/mickey_box_3.tex
\begin{tikzpicture}[scale=0.008]
\def\sx{1}
\def\sy{0.8}
\definecolor{R150G255B150}{RGB}{150,255,150};
\fill[R150G255B150] (200.000000*\sx, 231.440108*\sy) -- (200.000000*\sx, 427.739200*\sy) -- (370.000000*\sx, 427.739200*\sy) -- (370.000000*\sx, 231.440108*\sy) -- (200.000000*\sx, 231.440108*\sy) -- cycle;
\definecolor{black}{RGB}{0,0,0};
\draw[black] (200.000000*\sx, 231.440108*\sy) -- (200.000000*\sx, 427.739200*\sy) -- (370.000000*\sx, 427.739200*\sy) -- (370.000000*\sx, 231.440108*\sy) -- (200.000000*\sx, 231.440108*\sy) -- cycle;
\fill[R150G255B150] (200.000000*\sx, 35.141017*\sy) -- (200.000000*\sx, 231.440108*\sy) -- (370.000000*\sx, 231.440108*\sy) -- (370.000000*\sx, 35.141017*\sy) -- (200.000000*\sx, 35.141017*\sy) -- cycle;
\draw[black] (200.000000*\sx, 35.141017*\sy) -- (200.000000*\sx, 231.440108*\sy) -- (370.000000*\sx, 231.440108*\sy) -- (370.000000*\sx, 35.141017*\sy) -- (200.000000*\sx, 35.141017*\sy) -- cycle;
\fill[R150G255B150] (30.000000*\sx, 231.440108*\sy) -- (30.000000*\sx, 391.717645*\sy) -- (200.000000*\sx, 391.717645*\sy) -- (200.000000*\sx, 231.440108*\sy) -- (30.000000*\sx, 231.440108*\sy) -- cycle;
\draw[black] (30.000000*\sx, 231.440108*\sy) -- (30.000000*\sx, 391.717645*\sy) -- (200.000000*\sx, 391.717645*\sy) -- (200.000000*\sx, 231.440108*\sy) -- (30.000000*\sx, 231.440108*\sy) -- cycle;
\fill[R150G255B150] (30.000000*\sx, 71.162571*\sy) -- (30.000000*\sx, 231.440108*\sy) -- (200.000000*\sx, 231.440108*\sy) -- (200.000000*\sx, 71.162571*\sy) -- (30.000000*\sx, 71.162571*\sy) -- cycle;
\draw[black] (30.000000*\sx, 71.162571*\sy) -- (30.000000*\sx, 231.440108*\sy) -- (200.000000*\sx, 231.440108*\sy) -- (200.000000*\sx, 71.162571*\sy) -- (30.000000*\sx, 71.162571*\sy) -- cycle;
\definecolor{gray230}{RGB}{230,230,230};
\definecolor{gray128}{RGB}{128,128,128};
\fill[gray128] (400.000000*\sx, 8.005554*\sy) -- (400.000000*\sx, 4.541452*\sy) -- (0.000000*\sx, 4.541452*\sy) -- (0.000000*\sx, 8.005554*\sy) -- cycle;
\fill[gray128] (6.500000*\sx, 462.380216*\sy) -- (3.500000*\sx, 462.380216*\sy) -- (3.500000*\sx, 0.500000*\sy) -- (6.500000*\sx, 0.500000*\sy) -- cycle;
\draw[gray128] (30.000000*\sx,2.809402*\sy) -- (30.000000*\sx, 9.737605*\sy);
\definecolor{gray64}{RGB}{64,64,64};
\node[align=center,text=gray64,font=\tiny] at (30.000000*\sx, 17.820509*\sy) { 0 };
\draw[gray128] (115.000000*\sx,2.809402*\sy) -- (115.000000*\sx, 9.737605*\sy);
\node[align=center,text=gray64,font=\tiny] at (115.000000*\sx, 17.820509*\sy) { 0.5 };
\draw[gray128] (200.000000*\sx,2.809402*\sy) -- (200.000000*\sx, 9.737605*\sy);
\node[align=center,text=gray64,font=\tiny] at (200.000000*\sx, 17.820509*\sy) { 1 };
\draw[gray128] (285.000000*\sx,2.809402*\sy) -- (285.000000*\sx, 9.737605*\sy);
\node[align=center,text=gray64,font=\tiny] at (285.000000*\sx, 17.820509*\sy) { 1.5 };
\draw[gray128] (370.000000*\sx,2.809402*\sy) -- (370.000000*\sx, 9.737605*\sy);
\node[align=center,text=gray64,font=\tiny] at (370.000000*\sx, 17.820509*\sy) { 2 };
\draw[gray128] (455.000000*\sx,2.809402*\sy) -- (455.000000*\sx, 9.737605*\sy);
\node[align=center,text=gray64,font=\tiny] at (455.000000*\sx, 17.820509*\sy) { 2.5 };
\draw[gray128] (2.000000*\sx,4.773442*\sy) -- (8.000000*\sx, 4.773442*\sy);
\node[align=center,text=gray64,font=\tiny] at (5.000000*\sx, 4.773442*\sy) { -1 };
\draw[gray128] (2.000000*\sx,118.106775*\sy) -- (8.000000*\sx, 118.106775*\sy);
\node[align=center,text=gray64,font=\tiny] at (5.000000*\sx, 118.106775*\sy) { -0.5 };
\draw[gray128] (2.000000*\sx,231.440108*\sy) -- (8.000000*\sx, 231.440108*\sy);
\node[align=center,text=gray64,font=\tiny] at (5.000000*\sx, 231.440108*\sy) { 0 };
\draw[gray128] (2.000000*\sx,344.773442*\sy) -- (8.000000*\sx, 344.773442*\sy);
\node[align=center,text=gray64,font=\tiny] at (5.000000*\sx, 344.773442*\sy) { 0.5 };
\draw[gray128] (2.000000*\sx,458.106775*\sy) -- (8.000000*\sx, 458.106775*\sy);
\node[align=center,text=gray64,font=\tiny] at (5.000000*\sx, 458.106775*\sy) { 1 };
\node[align=center,text=black,font=\tiny] at (200.000000*\sx, 0.500000*\sy) { x };
\node[align=center,text=black,font=\tiny] at (0.000000*\sx, 231.440108*\sy) { y };
\end{tikzpicture}

%% file: imgs/mickey_box_5.tex
\begin{tikzpicture}[scale=0.008]
\def\sx{1}
\def\sy{0.8}
\definecolor{R150G255B150}{RGB}{150,255,150};
\definecolor{R150G150B255}{RGB}{150,150,255};
\fill[R150G255B150] (285.000000*\sx, 306.403062*\sy) -- (285.000000*\sx, 381.366016*\sy) -- (350.867964*\sx, 381.366016*\sy) -- (350.867964*\sx, 306.403062*\sy) -- (285.000000*\sx, 306.403062*\sy) -- cycle;
\definecolor{black}{RGB}{0,0,0};
\draw[black] (285.000000*\sx, 306.403062*\sy) -- (285.000000*\sx, 381.366016*\sy) -- (350.867964*\sx, 381.366016*\sy) -- (350.867964*\sx, 306.403062*\sy) -- (285.000000*\sx, 306.403062*\sy) -- cycle;
\fill[R150G255B150] (285.000000*\sx, 231.440108*\sy) -- (285.000000*\sx, 306.403062*\sy) -- (370.000000*\sx, 306.403062*\sy) -- (370.000000*\sx, 231.440108*\sy) -- (285.000000*\sx, 231.440108*\sy) -- cycle;
\draw[black] (285.000000*\sx, 231.440108*\sy) -- (285.000000*\sx, 306.403062*\sy) -- (370.000000*\sx, 306.403062*\sy) -- (370.000000*\sx, 231.440108*\sy) -- (285.000000*\sx, 231.440108*\sy) -- cycle;
\fill[R150G255B150] (200.000000*\sx, 329.589654*\sy) -- (200.000000*\sx, 427.739200*\sy) -- (285.000000*\sx, 427.739200*\sy) -- (285.000000*\sx, 329.589654*\sy) -- (200.000000*\sx, 329.589654*\sy) -- cycle;
\draw[black] (200.000000*\sx, 329.589654*\sy) -- (200.000000*\sx, 427.739200*\sy) -- (285.000000*\sx, 427.739200*\sy) -- (285.000000*\sx, 329.589654*\sy) -- (200.000000*\sx, 329.589654*\sy) -- cycle;
\fill[R150G255B150] (285.000000*\sx, 156.477154*\sy) -- (285.000000*\sx, 231.440108*\sy) -- (370.000000*\sx, 231.440108*\sy) -- (370.000000*\sx, 156.477154*\sy) -- (285.000000*\sx, 156.477154*\sy) -- cycle;
\draw[black] (285.000000*\sx, 156.477154*\sy) -- (285.000000*\sx, 231.440108*\sy) -- (370.000000*\sx, 231.440108*\sy) -- (370.000000*\sx, 156.477154*\sy) -- (285.000000*\sx, 156.477154*\sy) -- cycle;
\fill[R150G255B150] (285.000000*\sx, 81.514201*\sy) -- (285.000000*\sx, 156.477154*\sy) -- (350.867964*\sx, 156.477154*\sy) -- (350.867964*\sx, 81.514201*\sy) -- (285.000000*\sx, 81.514201*\sy) -- cycle;
\draw[black] (285.000000*\sx, 81.514201*\sy) -- (285.000000*\sx, 156.477154*\sy) -- (350.867964*\sx, 156.477154*\sy) -- (350.867964*\sx, 81.514201*\sy) -- (285.000000*\sx, 81.514201*\sy) -- cycle;
\fill[R150G255B150] (200.000000*\sx, 35.141017*\sy) -- (200.000000*\sx, 133.290562*\sy) -- (285.000000*\sx, 133.290562*\sy) -- (285.000000*\sx, 35.141017*\sy) -- (200.000000*\sx, 35.141017*\sy) -- cycle;
\draw[black] (200.000000*\sx, 35.141017*\sy) -- (200.000000*\sx, 133.290562*\sy) -- (285.000000*\sx, 133.290562*\sy) -- (285.000000*\sx, 35.141017*\sy) -- (200.000000*\sx, 35.141017*\sy) -- cycle;
\fill[R150G255B150] (115.000000*\sx, 311.578877*\sy) -- (115.000000*\sx, 391.717645*\sy) -- (200.000000*\sx, 391.717645*\sy) -- (200.000000*\sx, 311.578877*\sy) -- (115.000000*\sx, 311.578877*\sy) -- cycle;
\draw[black] (115.000000*\sx, 311.578877*\sy) -- (115.000000*\sx, 391.717645*\sy) -- (200.000000*\sx, 391.717645*\sy) -- (200.000000*\sx, 311.578877*\sy) -- (115.000000*\sx, 311.578877*\sy) -- cycle;
\fill[R150G255B150] (72.500000*\sx, 231.440108*\sy) -- (72.500000*\sx, 344.773442*\sy) -- (115.000000*\sx, 344.773442*\sy) -- (115.000000*\sx, 231.440108*\sy) -- (72.500000*\sx, 231.440108*\sy) -- cycle;
\draw[black] (72.500000*\sx, 231.440108*\sy) -- (72.500000*\sx, 344.773442*\sy) -- (115.000000*\sx, 344.773442*\sy) -- (115.000000*\sx, 231.440108*\sy) -- (72.500000*\sx, 231.440108*\sy) -- cycle;
\fill[R150G255B150] (30.000000*\sx, 231.440108*\sy) -- (30.000000*\sx, 311.578877*\sy) -- (72.500000*\sx, 311.578877*\sy) -- (72.500000*\sx, 231.440108*\sy) -- (30.000000*\sx, 231.440108*\sy) -- cycle;
\draw[black] (30.000000*\sx, 231.440108*\sy) -- (30.000000*\sx, 311.578877*\sy) -- (72.500000*\sx, 311.578877*\sy) -- (72.500000*\sx, 231.440108*\sy) -- (30.000000*\sx, 231.440108*\sy) -- cycle;
\fill[R150G255B150] (115.000000*\sx, 71.162571*\sy) -- (115.000000*\sx, 151.301340*\sy) -- (200.000000*\sx, 151.301340*\sy) -- (200.000000*\sx, 71.162571*\sy) -- (115.000000*\sx, 71.162571*\sy) -- cycle;
\draw[black] (115.000000*\sx, 71.162571*\sy) -- (115.000000*\sx, 151.301340*\sy) -- (200.000000*\sx, 151.301340*\sy) -- (200.000000*\sx, 71.162571*\sy) -- (115.000000*\sx, 71.162571*\sy) -- cycle;
\fill[R150G255B150] (72.500000*\sx, 118.106775*\sy) -- (72.500000*\sx, 231.440108*\sy) -- (115.000000*\sx, 231.440108*\sy) -- (115.000000*\sx, 118.106775*\sy) -- (72.500000*\sx, 118.106775*\sy) -- cycle;
\draw[black] (72.500000*\sx, 118.106775*\sy) -- (72.500000*\sx, 231.440108*\sy) -- (115.000000*\sx, 231.440108*\sy) -- (115.000000*\sx, 118.106775*\sy) -- (72.500000*\sx, 118.106775*\sy) -- cycle;
\fill[R150G255B150] (30.000000*\sx, 151.301340*\sy) -- (30.000000*\sx, 231.440108*\sy) -- (72.500000*\sx, 231.440108*\sy) -- (72.500000*\sx, 151.301340*\sy) -- (30.000000*\sx, 151.301340*\sy) -- cycle;
\draw[black] (30.000000*\sx, 151.301340*\sy) -- (30.000000*\sx, 231.440108*\sy) -- (72.500000*\sx, 231.440108*\sy) -- (72.500000*\sx, 151.301340*\sy) -- (30.000000*\sx, 151.301340*\sy) -- cycle;
\fill[R150G150B255] (200.000000*\sx, 231.440108*\sy) -- (200.000000*\sx, 329.589654*\sy) -- (285.000000*\sx, 329.589654*\sy) -- (285.000000*\sx, 231.440108*\sy) -- (200.000000*\sx, 231.440108*\sy) -- cycle;
\draw[black] (200.000000*\sx, 231.440108*\sy) -- (200.000000*\sx, 329.589654*\sy) -- (285.000000*\sx, 329.589654*\sy) -- (285.000000*\sx, 231.440108*\sy) -- (200.000000*\sx, 231.440108*\sy) -- cycle;
\fill[R150G150B255] (200.000000*\sx, 133.290562*\sy) -- (200.000000*\sx, 231.440108*\sy) -- (285.000000*\sx, 231.440108*\sy) -- (285.000000*\sx, 133.290562*\sy) -- (200.000000*\sx, 133.290562*\sy) -- cycle;
\draw[black] (200.000000*\sx, 133.290562*\sy) -- (200.000000*\sx, 231.440108*\sy) -- (285.000000*\sx, 231.440108*\sy) -- (285.000000*\sx, 133.290562*\sy) -- (200.000000*\sx, 133.290562*\sy) -- cycle;
\fill[R150G150B255] (115.000000*\sx, 231.440108*\sy) -- (115.000000*\sx, 311.578877*\sy) -- (200.000000*\sx, 311.578877*\sy) -- (200.000000*\sx, 231.440108*\sy) -- (115.000000*\sx, 231.440108*\sy) -- cycle;
\draw[black] (115.000000*\sx, 231.440108*\sy) -- (115.000000*\sx, 311.578877*\sy) -- (200.000000*\sx, 311.578877*\sy) -- (200.000000*\sx, 231.440108*\sy) -- (115.000000*\sx, 231.440108*\sy) -- cycle;
\fill[R150G150B255] (115.000000*\sx, 151.301340*\sy) -- (115.000000*\sx, 231.440108*\sy) -- (200.000000*\sx, 231.440108*\sy) -- (200.000000*\sx, 151.301340*\sy) -- (115.000000*\sx, 151.301340*\sy) -- cycle;
\draw[black] (115.000000*\sx, 151.301340*\sy) -- (115.000000*\sx, 231.440108*\sy) -- (200.000000*\sx, 231.440108*\sy) -- (200.000000*\sx, 151.301340*\sy) -- (115.000000*\sx, 151.301340*\sy) -- cycle;
\definecolor{gray230}{RGB}{230,230,230};
\definecolor{gray128}{RGB}{128,128,128};
\fill[gray128] (400.000000*\sx, 8.005554*\sy) -- (400.000000*\sx, 4.541452*\sy) -- (0.000000*\sx, 4.541452*\sy) -- (0.000000*\sx, 8.005554*\sy) -- cycle;
\fill[gray128] (6.500000*\sx, 462.380216*\sy) -- (3.500000*\sx, 462.380216*\sy) -- (3.500000*\sx, 0.500000*\sy) -- (6.500000*\sx, 0.500000*\sy) -- cycle;
\draw[gray128] (30.000000*\sx,2.809402*\sy) -- (30.000000*\sx, 9.737605*\sy);
\definecolor{gray64}{RGB}{64,64,64};
\node[align=center,text=gray64,font=\tiny] at (30.000000*\sx, 17.820509*\sy) { 0 };
\draw[gray128] (115.000000*\sx,2.809402*\sy) -- (115.000000*\sx, 9.737605*\sy);
\node[align=center,text=gray64,font=\tiny] at (115.000000*\sx, 17.820509*\sy) { 0.5 };
\draw[gray128] (200.000000*\sx,2.809402*\sy) -- (200.000000*\sx, 9.737605*\sy);
\node[align=center,text=gray64,font=\tiny] at (200.000000*\sx, 17.820509*\sy) { 1 };
\draw[gray128] (285.000000*\sx,2.809402*\sy) -- (285.000000*\sx, 9.737605*\sy);
\node[align=center,text=gray64,font=\tiny] at (285.000000*\sx, 17.820509*\sy) { 1.5 };
\draw[gray128] (370.000000*\sx,2.809402*\sy) -- (370.000000*\sx, 9.737605*\sy);
\node[align=center,text=gray64,font=\tiny] at (370.000000*\sx, 17.820509*\sy) { 2 };
\draw[gray128] (455.000000*\sx,2.809402*\sy) -- (455.000000*\sx, 9.737605*\sy);
\node[align=center,text=gray64,font=\tiny] at (455.000000*\sx, 17.820509*\sy) { 2.5 };
\draw[gray128] (2.000000*\sx,4.773442*\sy) -- (8.000000*\sx, 4.773442*\sy);
\node[align=center,text=gray64,font=\tiny] at (5.000000*\sx, 4.773442*\sy) { -1 };
\draw[gray128] (2.000000*\sx,118.106775*\sy) -- (8.000000*\sx, 118.106775*\sy);
\node[align=center,text=gray64,font=\tiny] at (5.000000*\sx, 118.106775*\sy) { -0.5 };
\draw[gray128] (2.000000*\sx,231.440108*\sy) -- (8.000000*\sx, 231.440108*\sy);
\node[align=center,text=gray64,font=\tiny] at (5.000000*\sx, 231.440108*\sy) { 0 };
\draw[gray128] (2.000000*\sx,344.773442*\sy) -- (8.000000*\sx, 344.773442*\sy);
\node[align=center,text=gray64,font=\tiny] at (5.000000*\sx, 344.773442*\sy) { 0.5 };
\draw[gray128] (2.000000*\sx,458.106775*\sy) -- (8.000000*\sx, 458.106775*\sy);
\node[align=center,text=gray64,font=\tiny] at (5.000000*\sx, 458.106775*\sy) { 1 };
\node[align=center,text=black,font=\tiny] at (200.000000*\sx, 0.500000*\sy) { x };
\node[align=center,text=black,font=\tiny] at (0.000000*\sx, 231.440108*\sy) { y };
\end{tikzpicture}

%% file: imgs/mickey_box_8.tex
\begin{tikzpicture}[scale=0.008]
\def\sx{1}
\def\sy{0.8}
\definecolor{R150G255B150}{RGB}{150,255,150};
\definecolor{R150G150B255}{RGB}{150,150,255};
\fill[R150G255B150] (327.345399*\sx, 370.785939*\sy) -- (327.345399*\sx, 385.386940*\sy) -- (336.756816*\sx, 385.386940*\sy) -- (336.756816*\sx, 370.785939*\sy) -- (327.345399*\sx, 370.785939*\sy) -- cycle;
\definecolor{black}{RGB}{0,0,0};
\draw[black] (327.345399*\sx, 370.785939*\sy) -- (327.345399*\sx, 385.386940*\sy) -- (336.756816*\sx, 385.386940*\sy) -- (336.756816*\sx, 370.785939*\sy) -- (327.345399*\sx, 370.785939*\sy) -- cycle;
\fill[R150G255B150] (317.933982*\sx, 370.785939*\sy) -- (317.933982*\sx, 398.134120*\sy) -- (327.345399*\sx, 398.134120*\sy) -- (327.345399*\sx, 370.785939*\sy) -- (317.933982*\sx, 370.785939*\sy) -- cycle;
\draw[black] (317.933982*\sx, 370.785939*\sy) -- (317.933982*\sx, 398.134120*\sy) -- (327.345399*\sx, 398.134120*\sy) -- (327.345399*\sx, 370.785939*\sy) -- (317.933982*\sx, 370.785939*\sy) -- cycle;
\fill[R150G255B150] (334.400973*\sx, 343.437758*\sy) -- (334.400973*\sx, 370.785939*\sy) -- (350.867964*\sx, 370.785939*\sy) -- (350.867964*\sx, 343.437758*\sy) -- (334.400973*\sx, 343.437758*\sy) -- cycle;
\draw[black] (334.400973*\sx, 343.437758*\sy) -- (334.400973*\sx, 370.785939*\sy) -- (350.867964*\sx, 370.785939*\sy) -- (350.867964*\sx, 343.437758*\sy) -- (334.400973*\sx, 343.437758*\sy) -- cycle;
\fill[R150G255B150] (301.466991*\sx, 388.415530*\sy) -- (301.466991*\sx, 417.249214*\sy) -- (317.933982*\sx, 417.249214*\sy) -- (317.933982*\sx, 388.415530*\sy) -- (301.466991*\sx, 388.415530*\sy) -- cycle;
\draw[black] (301.466991*\sx, 388.415530*\sy) -- (301.466991*\sx, 417.249214*\sy) -- (317.933982*\sx, 417.249214*\sy) -- (317.933982*\sx, 388.415530*\sy) -- (301.466991*\sx, 388.415530*\sy) -- cycle;
\fill[R150G255B150] (285.000000*\sx, 388.415530*\sy) -- (285.000000*\sx, 433.393302*\sy) -- (301.466991*\sx, 433.393302*\sy) -- (301.466991*\sx, 388.415530*\sy) -- (285.000000*\sx, 388.415530*\sy) -- cycle;
\draw[black] (285.000000*\sx, 388.415530*\sy) -- (285.000000*\sx, 433.393302*\sy) -- (301.466991*\sx, 433.393302*\sy) -- (301.466991*\sx, 388.415530*\sy) -- (285.000000*\sx, 388.415530*\sy) -- cycle;
\fill[R150G255B150] (346.409673*\sx, 298.459986*\sy) -- (346.409673*\sx, 343.437758*\sy) -- (365.319345*\sx, 343.437758*\sy) -- (365.319345*\sx, 298.459986*\sy) -- (346.409673*\sx, 298.459986*\sy) -- cycle;
\draw[black] (346.409673*\sx, 298.459986*\sy) -- (346.409673*\sx, 343.437758*\sy) -- (365.319345*\sx, 343.437758*\sy) -- (365.319345*\sx, 298.459986*\sy) -- (346.409673*\sx, 298.459986*\sy) -- cycle;
\fill[R150G255B150] (348.750000*\sx, 253.482213*\sy) -- (348.750000*\sx, 298.459986*\sy) -- (370.000000*\sx, 298.459986*\sy) -- (370.000000*\sx, 253.482213*\sy) -- (348.750000*\sx, 253.482213*\sy) -- cycle;
\draw[black] (348.750000*\sx, 253.482213*\sy) -- (348.750000*\sx, 298.459986*\sy) -- (370.000000*\sx, 298.459986*\sy) -- (370.000000*\sx, 253.482213*\sy) -- (348.750000*\sx, 253.482213*\sy) -- cycle;
\fill[R150G255B150] (263.750000*\sx, 418.536907*\sy) -- (263.750000*\sx, 451.004364*\sy) -- (285.000000*\sx, 451.004364*\sy) -- (285.000000*\sx, 418.536907*\sy) -- (263.750000*\sx, 418.536907*\sy) -- cycle;
\draw[black] (263.750000*\sx, 418.536907*\sy) -- (263.750000*\sx, 451.004364*\sy) -- (285.000000*\sx, 451.004364*\sy) -- (285.000000*\sx, 418.536907*\sy) -- (263.750000*\sx, 418.536907*\sy) -- cycle;
\fill[R150G255B150] (242.500000*\sx, 418.536907*\sy) -- (242.500000*\sx, 465.812145*\sy) -- (263.750000*\sx, 465.812145*\sy) -- (263.750000*\sx, 418.536907*\sy) -- (242.500000*\sx, 418.536907*\sy) -- cycle;
\draw[black] (242.500000*\sx, 418.536907*\sy) -- (242.500000*\sx, 465.812145*\sy) -- (263.750000*\sx, 465.812145*\sy) -- (263.750000*\sx, 418.536907*\sy) -- (242.500000*\sx, 418.536907*\sy) -- cycle;
\fill[R150G255B150] (221.250000*\sx, 419.889381*\sy) -- (221.250000*\sx, 468.517094*\sy) -- (242.500000*\sx, 468.517094*\sy) -- (242.500000*\sx, 419.889381*\sy) -- (221.250000*\sx, 419.889381*\sy) -- cycle;
\draw[black] (221.250000*\sx, 419.889381*\sy) -- (221.250000*\sx, 468.517094*\sy) -- (242.500000*\sx, 468.517094*\sy) -- (242.500000*\sx, 419.889381*\sy) -- (221.250000*\sx, 419.889381*\sy) -- cycle;
\fill[R150G255B150] (200.000000*\sx, 419.889381*\sy) -- (200.000000*\sx, 457.482213*\sy) -- (221.250000*\sx, 457.482213*\sy) -- (221.250000*\sx, 419.889381*\sy) -- (200.000000*\sx, 419.889381*\sy) -- cycle;
\draw[black] (200.000000*\sx, 419.889381*\sy) -- (200.000000*\sx, 457.482213*\sy) -- (221.250000*\sx, 457.482213*\sy) -- (221.250000*\sx, 419.889381*\sy) -- (200.000000*\sx, 419.889381*\sy) -- cycle;
\fill[R150G255B150] (348.750000*\sx, 208.504441*\sy) -- (348.750000*\sx, 253.482213*\sy) -- (370.000000*\sx, 253.482213*\sy) -- (370.000000*\sx, 208.504441*\sy) -- (348.750000*\sx, 208.504441*\sy) -- cycle;
\draw[black] (348.750000*\sx, 208.504441*\sy) -- (348.750000*\sx, 253.482213*\sy) -- (370.000000*\sx, 253.482213*\sy) -- (370.000000*\sx, 208.504441*\sy) -- (348.750000*\sx, 208.504441*\sy) -- cycle;
\fill[R150G255B150] (346.409673*\sx, 163.526669*\sy) -- (346.409673*\sx, 208.504441*\sy) -- (365.319345*\sx, 208.504441*\sy) -- (365.319345*\sx, 163.526669*\sy) -- (346.409673*\sx, 163.526669*\sy) -- cycle;
\draw[black] (346.409673*\sx, 163.526669*\sy) -- (346.409673*\sx, 208.504441*\sy) -- (365.319345*\sx, 208.504441*\sy) -- (365.319345*\sx, 163.526669*\sy) -- (346.409673*\sx, 163.526669*\sy) -- cycle;
\fill[R150G255B150] (334.400973*\sx, 136.178488*\sy) -- (334.400973*\sx, 163.526669*\sy) -- (350.867964*\sx, 163.526669*\sy) -- (350.867964*\sx, 136.178488*\sy) -- (334.400973*\sx, 136.178488*\sy) -- cycle;
\draw[black] (334.400973*\sx, 136.178488*\sy) -- (334.400973*\sx, 163.526669*\sy) -- (350.867964*\sx, 163.526669*\sy) -- (350.867964*\sx, 136.178488*\sy) -- (334.400973*\sx, 136.178488*\sy) -- cycle;
\fill[R150G255B150] (327.345399*\sx, 121.577486*\sy) -- (327.345399*\sx, 136.178488*\sy) -- (336.756816*\sx, 136.178488*\sy) -- (336.756816*\sx, 121.577486*\sy) -- (327.345399*\sx, 121.577486*\sy) -- cycle;
\draw[black] (327.345399*\sx, 121.577486*\sy) -- (327.345399*\sx, 136.178488*\sy) -- (336.756816*\sx, 136.178488*\sy) -- (336.756816*\sx, 121.577486*\sy) -- (327.345399*\sx, 121.577486*\sy) -- cycle;
\fill[R150G255B150] (317.933982*\sx, 108.830306*\sy) -- (317.933982*\sx, 136.178488*\sy) -- (327.345399*\sx, 136.178488*\sy) -- (327.345399*\sx, 108.830306*\sy) -- (317.933982*\sx, 108.830306*\sy) -- cycle;
\draw[black] (317.933982*\sx, 108.830306*\sy) -- (317.933982*\sx, 136.178488*\sy) -- (327.345399*\sx, 136.178488*\sy) -- (327.345399*\sx, 108.830306*\sy) -- (317.933982*\sx, 108.830306*\sy) -- cycle;
\fill[R150G255B150] (301.466991*\sx, 89.715212*\sy) -- (301.466991*\sx, 118.548896*\sy) -- (317.933982*\sx, 118.548896*\sy) -- (317.933982*\sx, 89.715212*\sy) -- (301.466991*\sx, 89.715212*\sy) -- cycle;
\draw[black] (301.466991*\sx, 89.715212*\sy) -- (301.466991*\sx, 118.548896*\sy) -- (317.933982*\sx, 118.548896*\sy) -- (317.933982*\sx, 89.715212*\sy) -- (301.466991*\sx, 89.715212*\sy) -- cycle;
\fill[R150G255B150] (285.000000*\sx, 73.571124*\sy) -- (285.000000*\sx, 118.548896*\sy) -- (301.466991*\sx, 118.548896*\sy) -- (301.466991*\sx, 73.571124*\sy) -- (285.000000*\sx, 73.571124*\sy) -- cycle;
\draw[black] (285.000000*\sx, 73.571124*\sy) -- (285.000000*\sx, 118.548896*\sy) -- (301.466991*\sx, 118.548896*\sy) -- (301.466991*\sx, 73.571124*\sy) -- (285.000000*\sx, 73.571124*\sy) -- cycle;
\fill[R150G255B150] (263.750000*\sx, 55.960063*\sy) -- (263.750000*\sx, 88.427520*\sy) -- (285.000000*\sx, 88.427520*\sy) -- (285.000000*\sx, 55.960063*\sy) -- (263.750000*\sx, 55.960063*\sy) -- cycle;
\draw[black] (263.750000*\sx, 55.960063*\sy) -- (263.750000*\sx, 88.427520*\sy) -- (285.000000*\sx, 88.427520*\sy) -- (285.000000*\sx, 55.960063*\sy) -- (263.750000*\sx, 55.960063*\sy) -- cycle;
\fill[R150G255B150] (242.500000*\sx, 41.152281*\sy) -- (242.500000*\sx, 88.427520*\sy) -- (263.750000*\sx, 88.427520*\sy) -- (263.750000*\sx, 41.152281*\sy) -- (242.500000*\sx, 41.152281*\sy) -- cycle;
\draw[black] (242.500000*\sx, 41.152281*\sy) -- (242.500000*\sx, 88.427520*\sy) -- (263.750000*\sx, 88.427520*\sy) -- (263.750000*\sx, 41.152281*\sy) -- (242.500000*\sx, 41.152281*\sy) -- cycle;
\fill[R150G255B150] (221.250000*\sx, 38.447332*\sy) -- (221.250000*\sx, 87.075045*\sy) -- (242.500000*\sx, 87.075045*\sy) -- (242.500000*\sx, 38.447332*\sy) -- (221.250000*\sx, 38.447332*\sy) -- cycle;
\draw[black] (221.250000*\sx, 38.447332*\sy) -- (221.250000*\sx, 87.075045*\sy) -- (242.500000*\sx, 87.075045*\sy) -- (242.500000*\sx, 38.447332*\sy) -- (221.250000*\sx, 38.447332*\sy) -- cycle;
\fill[R150G255B150] (200.000000*\sx, 49.482213*\sy) -- (200.000000*\sx, 87.075045*\sy) -- (221.250000*\sx, 87.075045*\sy) -- (221.250000*\sx, 49.482213*\sy) -- (200.000000*\sx, 49.482213*\sy) -- cycle;
\draw[black] (200.000000*\sx, 49.482213*\sy) -- (200.000000*\sx, 87.075045*\sy) -- (221.250000*\sx, 87.075045*\sy) -- (221.250000*\sx, 49.482213*\sy) -- (200.000000*\sx, 49.482213*\sy) -- cycle;
\fill[R150G255B150] (178.750000*\sx, 397.731997*\sy) -- (178.750000*\sx, 445.815258*\sy) -- (200.000000*\sx, 445.815258*\sy) -- (200.000000*\sx, 397.731997*\sy) -- (178.750000*\sx, 397.731997*\sy) -- cycle;
\draw[black] (178.750000*\sx, 397.731997*\sy) -- (178.750000*\sx, 445.815258*\sy) -- (200.000000*\sx, 445.815258*\sy) -- (200.000000*\sx, 397.731997*\sy) -- (178.750000*\sx, 397.731997*\sy) -- cycle;
\fill[R150G255B150] (157.500000*\sx, 397.731997*\sy) -- (157.500000*\sx, 433.393302*\sy) -- (178.750000*\sx, 433.393302*\sy) -- (178.750000*\sx, 397.731997*\sy) -- (157.500000*\sx, 397.731997*\sy) -- cycle;
\draw[black] (157.500000*\sx, 397.731997*\sy) -- (157.500000*\sx, 433.393302*\sy) -- (178.750000*\sx, 433.393302*\sy) -- (178.750000*\sx, 397.731997*\sy) -- (157.500000*\sx, 397.731997*\sy) -- cycle;
\fill[R150G255B150] (136.250000*\sx, 384.848126*\sy) -- (136.250000*\sx, 420.047516*\sy) -- (157.500000*\sx, 420.047516*\sy) -- (157.500000*\sx, 384.848126*\sy) -- (136.250000*\sx, 384.848126*\sy) -- cycle;
\draw[black] (136.250000*\sx, 384.848126*\sy) -- (136.250000*\sx, 420.047516*\sy) -- (157.500000*\sx, 420.047516*\sy) -- (157.500000*\sx, 384.848126*\sy) -- (136.250000*\sx, 384.848126*\sy) -- cycle;
\fill[R150G255B150] (115.000000*\sx, 384.848126*\sy) -- (115.000000*\sx, 405.534836*\sy) -- (136.250000*\sx, 405.534836*\sy) -- (136.250000*\sx, 384.848126*\sy) -- (115.000000*\sx, 384.848126*\sy) -- cycle;
\draw[black] (115.000000*\sx, 384.848126*\sy) -- (115.000000*\sx, 405.534836*\sy) -- (136.250000*\sx, 405.534836*\sy) -- (136.250000*\sx, 384.848126*\sy) -- (115.000000*\sx, 384.848126*\sy) -- cycle;
\fill[R150G255B150] (93.750000*\sx, 355.482213*\sy) -- (93.750000*\sx, 389.482213*\sy) -- (115.000000*\sx, 389.482213*\sy) -- (115.000000*\sx, 355.482213*\sy) -- (93.750000*\sx, 355.482213*\sy) -- cycle;
\draw[black] (93.750000*\sx, 355.482213*\sy) -- (93.750000*\sx, 389.482213*\sy) -- (115.000000*\sx, 389.482213*\sy) -- (115.000000*\sx, 355.482213*\sy) -- (93.750000*\sx, 355.482213*\sy) -- cycle;
\fill[R150G255B150] (72.500000*\sx, 346.371941*\sy) -- (72.500000*\sx, 371.261668*\sy) -- (93.750000*\sx, 371.261668*\sy) -- (93.750000*\sx, 346.371941*\sy) -- (72.500000*\sx, 346.371941*\sy) -- cycle;
\draw[black] (72.500000*\sx, 346.371941*\sy) -- (72.500000*\sx, 371.261668*\sy) -- (93.750000*\sx, 371.261668*\sy) -- (93.750000*\sx, 346.371941*\sy) -- (72.500000*\sx, 346.371941*\sy) -- cycle;
\fill[R150G255B150] (56.562500*\sx, 325.607105*\sy) -- (56.562500*\sx, 349.648736*\sy) -- (72.500000*\sx, 349.648736*\sy) -- (72.500000*\sx, 325.607105*\sy) -- (56.562500*\sx, 325.607105*\sy) -- cycle;
\draw[black] (56.562500*\sx, 325.607105*\sy) -- (56.562500*\sx, 349.648736*\sy) -- (72.500000*\sx, 349.648736*\sy) -- (72.500000*\sx, 325.607105*\sy) -- (56.562500*\sx, 325.607105*\sy) -- cycle;
\fill[R150G255B150] (47.696676*\sx, 315.536999*\sy) -- (47.696676*\sx, 329.508525*\sy) -- (56.562500*\sx, 329.508525*\sy) -- (56.562500*\sx, 315.536999*\sy) -- (47.696676*\sx, 315.536999*\sy) -- cycle;
\draw[black] (47.696676*\sx, 315.536999*\sy) -- (47.696676*\sx, 329.508525*\sy) -- (56.562500*\sx, 329.508525*\sy) -- (56.562500*\sx, 315.536999*\sy) -- (47.696676*\sx, 315.536999*\sy) -- cycle;
\fill[R150G255B150] (40.625000*\sx, 301.565474*\sy) -- (40.625000*\sx, 315.536999*\sy) -- (56.562500*\sx, 315.536999*\sy) -- (56.562500*\sx, 301.565474*\sy) -- (40.625000*\sx, 301.565474*\sy) -- cycle;
\draw[black] (40.625000*\sx, 301.565474*\sy) -- (40.625000*\sx, 315.536999*\sy) -- (56.562500*\sx, 315.536999*\sy) -- (56.562500*\sx, 301.565474*\sy) -- (40.625000*\sx, 301.565474*\sy) -- cycle;
\fill[R150G255B150] (32.656250*\sx, 277.523844*\sy) -- (32.656250*\sx, 301.565474*\sy) -- (51.250000*\sx, 301.565474*\sy) -- (51.250000*\sx, 277.523844*\sy) -- (32.656250*\sx, 277.523844*\sy) -- cycle;
\draw[black] (32.656250*\sx, 277.523844*\sy) -- (32.656250*\sx, 301.565474*\sy) -- (51.250000*\sx, 301.565474*\sy) -- (51.250000*\sx, 277.523844*\sy) -- (32.656250*\sx, 277.523844*\sy) -- cycle;
\fill[R150G255B150] (30.000000*\sx, 253.482213*\sy) -- (30.000000*\sx, 277.523844*\sy) -- (51.250000*\sx, 277.523844*\sy) -- (51.250000*\sx, 253.482213*\sy) -- (30.000000*\sx, 253.482213*\sy) -- cycle;
\draw[black] (30.000000*\sx, 253.482213*\sy) -- (30.000000*\sx, 277.523844*\sy) -- (51.250000*\sx, 277.523844*\sy) -- (51.250000*\sx, 253.482213*\sy) -- (30.000000*\sx, 253.482213*\sy) -- cycle;
\fill[R150G255B150] (178.750000*\sx, 61.149169*\sy) -- (178.750000*\sx, 109.232430*\sy) -- (200.000000*\sx, 109.232430*\sy) -- (200.000000*\sx, 61.149169*\sy) -- (178.750000*\sx, 61.149169*\sy) -- cycle;
\draw[black] (178.750000*\sx, 61.149169*\sy) -- (178.750000*\sx, 109.232430*\sy) -- (200.000000*\sx, 109.232430*\sy) -- (200.000000*\sx, 61.149169*\sy) -- (178.750000*\sx, 61.149169*\sy) -- cycle;
\fill[R150G255B150] (157.500000*\sx, 73.571124*\sy) -- (157.500000*\sx, 109.232430*\sy) -- (178.750000*\sx, 109.232430*\sy) -- (178.750000*\sx, 73.571124*\sy) -- (157.500000*\sx, 73.571124*\sy) -- cycle;
\draw[black] (157.500000*\sx, 73.571124*\sy) -- (157.500000*\sx, 109.232430*\sy) -- (178.750000*\sx, 109.232430*\sy) -- (178.750000*\sx, 73.571124*\sy) -- (157.500000*\sx, 73.571124*\sy) -- cycle;
\fill[R150G255B150] (136.250000*\sx, 86.916911*\sy) -- (136.250000*\sx, 122.116301*\sy) -- (157.500000*\sx, 122.116301*\sy) -- (157.500000*\sx, 86.916911*\sy) -- (136.250000*\sx, 86.916911*\sy) -- cycle;
\draw[black] (136.250000*\sx, 86.916911*\sy) -- (136.250000*\sx, 122.116301*\sy) -- (157.500000*\sx, 122.116301*\sy) -- (157.500000*\sx, 86.916911*\sy) -- (136.250000*\sx, 86.916911*\sy) -- cycle;
\fill[R150G255B150] (115.000000*\sx, 101.429591*\sy) -- (115.000000*\sx, 122.116301*\sy) -- (136.250000*\sx, 122.116301*\sy) -- (136.250000*\sx, 101.429591*\sy) -- (115.000000*\sx, 101.429591*\sy) -- cycle;
\draw[black] (115.000000*\sx, 101.429591*\sy) -- (115.000000*\sx, 122.116301*\sy) -- (136.250000*\sx, 122.116301*\sy) -- (136.250000*\sx, 101.429591*\sy) -- (115.000000*\sx, 101.429591*\sy) -- cycle;
\fill[R150G255B150] (93.750000*\sx, 117.482213*\sy) -- (93.750000*\sx, 151.482213*\sy) -- (115.000000*\sx, 151.482213*\sy) -- (115.000000*\sx, 117.482213*\sy) -- (93.750000*\sx, 117.482213*\sy) -- cycle;
\draw[black] (93.750000*\sx, 117.482213*\sy) -- (93.750000*\sx, 151.482213*\sy) -- (115.000000*\sx, 151.482213*\sy) -- (115.000000*\sx, 117.482213*\sy) -- (93.750000*\sx, 117.482213*\sy) -- cycle;
\fill[R150G255B150] (72.500000*\sx, 135.702758*\sy) -- (72.500000*\sx, 160.592486*\sy) -- (93.750000*\sx, 160.592486*\sy) -- (93.750000*\sx, 135.702758*\sy) -- (72.500000*\sx, 135.702758*\sy) -- cycle;
\draw[black] (72.500000*\sx, 135.702758*\sy) -- (72.500000*\sx, 160.592486*\sy) -- (93.750000*\sx, 160.592486*\sy) -- (93.750000*\sx, 135.702758*\sy) -- (72.500000*\sx, 135.702758*\sy) -- cycle;
\fill[R150G255B150] (30.000000*\sx, 229.440583*\sy) -- (30.000000*\sx, 253.482213*\sy) -- (51.250000*\sx, 253.482213*\sy) -- (51.250000*\sx, 229.440583*\sy) -- (30.000000*\sx, 229.440583*\sy) -- cycle;
\draw[black] (30.000000*\sx, 229.440583*\sy) -- (30.000000*\sx, 253.482213*\sy) -- (51.250000*\sx, 253.482213*\sy) -- (51.250000*\sx, 229.440583*\sy) -- (30.000000*\sx, 229.440583*\sy) -- cycle;
\fill[R150G255B150] (32.656250*\sx, 205.398952*\sy) -- (32.656250*\sx, 229.440583*\sy) -- (51.250000*\sx, 229.440583*\sy) -- (51.250000*\sx, 205.398952*\sy) -- (32.656250*\sx, 205.398952*\sy) -- cycle;
\draw[black] (32.656250*\sx, 205.398952*\sy) -- (32.656250*\sx, 229.440583*\sy) -- (51.250000*\sx, 229.440583*\sy) -- (51.250000*\sx, 205.398952*\sy) -- (32.656250*\sx, 205.398952*\sy) -- cycle;
\fill[R150G255B150] (56.562500*\sx, 157.315691*\sy) -- (56.562500*\sx, 181.357322*\sy) -- (72.500000*\sx, 181.357322*\sy) -- (72.500000*\sx, 157.315691*\sy) -- (56.562500*\sx, 157.315691*\sy) -- cycle;
\draw[black] (56.562500*\sx, 157.315691*\sy) -- (56.562500*\sx, 181.357322*\sy) -- (72.500000*\sx, 181.357322*\sy) -- (72.500000*\sx, 157.315691*\sy) -- (56.562500*\sx, 157.315691*\sy) -- cycle;
\fill[R150G255B150] (40.625000*\sx, 191.427427*\sy) -- (40.625000*\sx, 205.398952*\sy) -- (56.562500*\sx, 205.398952*\sy) -- (56.562500*\sx, 191.427427*\sy) -- (40.625000*\sx, 191.427427*\sy) -- cycle;
\draw[black] (40.625000*\sx, 191.427427*\sy) -- (40.625000*\sx, 205.398952*\sy) -- (56.562500*\sx, 205.398952*\sy) -- (56.562500*\sx, 191.427427*\sy) -- (40.625000*\sx, 191.427427*\sy) -- cycle;
\fill[R150G255B150] (47.696676*\sx, 177.455902*\sy) -- (47.696676*\sx, 191.427427*\sy) -- (56.562500*\sx, 191.427427*\sy) -- (56.562500*\sx, 177.455902*\sy) -- (47.696676*\sx, 177.455902*\sy) -- cycle;
\draw[black] (47.696676*\sx, 177.455902*\sy) -- (47.696676*\sx, 191.427427*\sy) -- (56.562500*\sx, 191.427427*\sy) -- (56.562500*\sx, 177.455902*\sy) -- (47.696676*\sx, 177.455902*\sy) -- cycle;
\fill[R150G150B255] (317.933982*\sx, 343.437758*\sy) -- (317.933982*\sx, 370.785939*\sy) -- (334.400973*\sx, 370.785939*\sy) -- (334.400973*\sx, 343.437758*\sy) -- (317.933982*\sx, 343.437758*\sy) -- cycle;
\draw[black] (317.933982*\sx, 343.437758*\sy) -- (317.933982*\sx, 370.785939*\sy) -- (334.400973*\sx, 370.785939*\sy) -- (334.400973*\sx, 343.437758*\sy) -- (317.933982*\sx, 343.437758*\sy) -- cycle;
\fill[R150G150B255] (285.000000*\sx, 343.437758*\sy) -- (285.000000*\sx, 388.415530*\sy) -- (317.933982*\sx, 388.415530*\sy) -- (317.933982*\sx, 343.437758*\sy) -- (285.000000*\sx, 343.437758*\sy) -- cycle;
\draw[black] (285.000000*\sx, 343.437758*\sy) -- (285.000000*\sx, 388.415530*\sy) -- (317.933982*\sx, 388.415530*\sy) -- (317.933982*\sx, 343.437758*\sy) -- (285.000000*\sx, 343.437758*\sy) -- cycle;
\fill[R150G150B255] (327.500000*\sx, 298.459986*\sy) -- (327.500000*\sx, 343.437758*\sy) -- (346.409673*\sx, 343.437758*\sy) -- (346.409673*\sx, 298.459986*\sy) -- (327.500000*\sx, 298.459986*\sy) -- cycle;
\draw[black] (327.500000*\sx, 298.459986*\sy) -- (327.500000*\sx, 343.437758*\sy) -- (346.409673*\sx, 343.437758*\sy) -- (346.409673*\sx, 298.459986*\sy) -- (327.500000*\sx, 298.459986*\sy) -- cycle;
\fill[R150G150B255] (327.500000*\sx, 253.482213*\sy) -- (327.500000*\sx, 298.459986*\sy) -- (348.750000*\sx, 298.459986*\sy) -- (348.750000*\sx, 253.482213*\sy) -- (327.500000*\sx, 253.482213*\sy) -- cycle;
\draw[black] (327.500000*\sx, 253.482213*\sy) -- (327.500000*\sx, 298.459986*\sy) -- (348.750000*\sx, 298.459986*\sy) -- (348.750000*\sx, 253.482213*\sy) -- (327.500000*\sx, 253.482213*\sy) -- cycle;
\fill[R150G150B255] (285.000000*\sx, 253.482213*\sy) -- (285.000000*\sx, 343.437758*\sy) -- (327.500000*\sx, 343.437758*\sy) -- (327.500000*\sx, 253.482213*\sy) -- (285.000000*\sx, 253.482213*\sy) -- cycle;
\draw[black] (285.000000*\sx, 253.482213*\sy) -- (285.000000*\sx, 343.437758*\sy) -- (327.500000*\sx, 343.437758*\sy) -- (327.500000*\sx, 253.482213*\sy) -- (285.000000*\sx, 253.482213*\sy) -- cycle;
\fill[R150G150B255] (242.500000*\sx, 371.261668*\sy) -- (242.500000*\sx, 418.536907*\sy) -- (285.000000*\sx, 418.536907*\sy) -- (285.000000*\sx, 371.261668*\sy) -- (242.500000*\sx, 371.261668*\sy) -- cycle;
\draw[black] (242.500000*\sx, 371.261668*\sy) -- (242.500000*\sx, 418.536907*\sy) -- (285.000000*\sx, 418.536907*\sy) -- (285.000000*\sx, 371.261668*\sy) -- (242.500000*\sx, 371.261668*\sy) -- cycle;
\fill[R150G150B255] (200.000000*\sx, 371.261668*\sy) -- (200.000000*\sx, 419.889381*\sy) -- (242.500000*\sx, 419.889381*\sy) -- (242.500000*\sx, 371.261668*\sy) -- (200.000000*\sx, 371.261668*\sy) -- cycle;
\draw[black] (200.000000*\sx, 371.261668*\sy) -- (200.000000*\sx, 419.889381*\sy) -- (242.500000*\sx, 419.889381*\sy) -- (242.500000*\sx, 371.261668*\sy) -- (200.000000*\sx, 371.261668*\sy) -- cycle;
\fill[R150G150B255] (200.000000*\sx, 253.482213*\sy) -- (200.000000*\sx, 371.261668*\sy) -- (285.000000*\sx, 371.261668*\sy) -- (285.000000*\sx, 253.482213*\sy) -- (200.000000*\sx, 253.482213*\sy) -- cycle;
\draw[black] (200.000000*\sx, 253.482213*\sy) -- (200.000000*\sx, 371.261668*\sy) -- (285.000000*\sx, 371.261668*\sy) -- (285.000000*\sx, 253.482213*\sy) -- (200.000000*\sx, 253.482213*\sy) -- cycle;
\fill[R150G150B255] (327.500000*\sx, 208.504441*\sy) -- (327.500000*\sx, 253.482213*\sy) -- (348.750000*\sx, 253.482213*\sy) -- (348.750000*\sx, 208.504441*\sy) -- (327.500000*\sx, 208.504441*\sy) -- cycle;
\draw[black] (327.500000*\sx, 208.504441*\sy) -- (327.500000*\sx, 253.482213*\sy) -- (348.750000*\sx, 253.482213*\sy) -- (348.750000*\sx, 208.504441*\sy) -- (327.500000*\sx, 208.504441*\sy) -- cycle;
\fill[R150G150B255] (327.500000*\sx, 163.526669*\sy) -- (327.500000*\sx, 208.504441*\sy) -- (346.409673*\sx, 208.504441*\sy) -- (346.409673*\sx, 163.526669*\sy) -- (327.500000*\sx, 163.526669*\sy) -- cycle;
\draw[black] (327.500000*\sx, 163.526669*\sy) -- (327.500000*\sx, 208.504441*\sy) -- (346.409673*\sx, 208.504441*\sy) -- (346.409673*\sx, 163.526669*\sy) -- (327.500000*\sx, 163.526669*\sy) -- cycle;
\fill[R150G150B255] (285.000000*\sx, 163.526669*\sy) -- (285.000000*\sx, 253.482213*\sy) -- (327.500000*\sx, 253.482213*\sy) -- (327.500000*\sx, 163.526669*\sy) -- (285.000000*\sx, 163.526669*\sy) -- cycle;
\draw[black] (285.000000*\sx, 163.526669*\sy) -- (285.000000*\sx, 253.482213*\sy) -- (327.500000*\sx, 253.482213*\sy) -- (327.500000*\sx, 163.526669*\sy) -- (285.000000*\sx, 163.526669*\sy) -- cycle;
\fill[R150G150B255] (317.933982*\sx, 136.178488*\sy) -- (317.933982*\sx, 163.526669*\sy) -- (334.400973*\sx, 163.526669*\sy) -- (334.400973*\sx, 136.178488*\sy) -- (317.933982*\sx, 136.178488*\sy) -- cycle;
\draw[black] (317.933982*\sx, 136.178488*\sy) -- (317.933982*\sx, 163.526669*\sy) -- (334.400973*\sx, 163.526669*\sy) -- (334.400973*\sx, 136.178488*\sy) -- (317.933982*\sx, 136.178488*\sy) -- cycle;
\fill[R150G150B255] (285.000000*\sx, 118.548896*\sy) -- (285.000000*\sx, 163.526669*\sy) -- (317.933982*\sx, 163.526669*\sy) -- (317.933982*\sx, 118.548896*\sy) -- (285.000000*\sx, 118.548896*\sy) -- cycle;
\draw[black] (285.000000*\sx, 118.548896*\sy) -- (285.000000*\sx, 163.526669*\sy) -- (317.933982*\sx, 163.526669*\sy) -- (317.933982*\sx, 118.548896*\sy) -- (285.000000*\sx, 118.548896*\sy) -- cycle;
\fill[R150G150B255] (200.000000*\sx, 135.702758*\sy) -- (200.000000*\sx, 253.482213*\sy) -- (285.000000*\sx, 253.482213*\sy) -- (285.000000*\sx, 135.702758*\sy) -- (200.000000*\sx, 135.702758*\sy) -- cycle;
\draw[black] (200.000000*\sx, 135.702758*\sy) -- (200.000000*\sx, 253.482213*\sy) -- (285.000000*\sx, 253.482213*\sy) -- (285.000000*\sx, 135.702758*\sy) -- (200.000000*\sx, 135.702758*\sy) -- cycle;
\fill[R150G150B255] (242.500000*\sx, 88.427520*\sy) -- (242.500000*\sx, 135.702758*\sy) -- (285.000000*\sx, 135.702758*\sy) -- (285.000000*\sx, 88.427520*\sy) -- (242.500000*\sx, 88.427520*\sy) -- cycle;
\draw[black] (242.500000*\sx, 88.427520*\sy) -- (242.500000*\sx, 135.702758*\sy) -- (285.000000*\sx, 135.702758*\sy) -- (285.000000*\sx, 88.427520*\sy) -- (242.500000*\sx, 88.427520*\sy) -- cycle;
\fill[R150G150B255] (200.000000*\sx, 87.075045*\sy) -- (200.000000*\sx, 135.702758*\sy) -- (242.500000*\sx, 135.702758*\sy) -- (242.500000*\sx, 87.075045*\sy) -- (200.000000*\sx, 87.075045*\sy) -- cycle;
\draw[black] (200.000000*\sx, 87.075045*\sy) -- (200.000000*\sx, 135.702758*\sy) -- (242.500000*\sx, 135.702758*\sy) -- (242.500000*\sx, 87.075045*\sy) -- (200.000000*\sx, 87.075045*\sy) -- cycle;
\fill[R150G150B255] (157.500000*\sx, 349.648736*\sy) -- (157.500000*\sx, 397.731997*\sy) -- (200.000000*\sx, 397.731997*\sy) -- (200.000000*\sx, 349.648736*\sy) -- (157.500000*\sx, 349.648736*\sy) -- cycle;
\draw[black] (157.500000*\sx, 349.648736*\sy) -- (157.500000*\sx, 397.731997*\sy) -- (200.000000*\sx, 397.731997*\sy) -- (200.000000*\sx, 349.648736*\sy) -- (157.500000*\sx, 349.648736*\sy) -- cycle;
\fill[R150G150B255] (115.000000*\sx, 349.648736*\sy) -- (115.000000*\sx, 384.848126*\sy) -- (157.500000*\sx, 384.848126*\sy) -- (157.500000*\sx, 349.648736*\sy) -- (115.000000*\sx, 349.648736*\sy) -- cycle;
\draw[black] (115.000000*\sx, 349.648736*\sy) -- (115.000000*\sx, 384.848126*\sy) -- (157.500000*\sx, 384.848126*\sy) -- (157.500000*\sx, 349.648736*\sy) -- (115.000000*\sx, 349.648736*\sy) -- cycle;
\fill[R150G150B255] (115.000000*\sx, 253.482213*\sy) -- (115.000000*\sx, 349.648736*\sy) -- (200.000000*\sx, 349.648736*\sy) -- (200.000000*\sx, 253.482213*\sy) -- (115.000000*\sx, 253.482213*\sy) -- cycle;
\draw[black] (115.000000*\sx, 253.482213*\sy) -- (115.000000*\sx, 349.648736*\sy) -- (200.000000*\sx, 349.648736*\sy) -- (200.000000*\sx, 253.482213*\sy) -- (115.000000*\sx, 253.482213*\sy) -- cycle;
\fill[R150G150B255] (93.750000*\sx, 321.482213*\sy) -- (93.750000*\sx, 355.482213*\sy) -- (115.000000*\sx, 355.482213*\sy) -- (115.000000*\sx, 321.482213*\sy) -- (93.750000*\sx, 321.482213*\sy) -- cycle;
\draw[black] (93.750000*\sx, 321.482213*\sy) -- (93.750000*\sx, 355.482213*\sy) -- (115.000000*\sx, 355.482213*\sy) -- (115.000000*\sx, 321.482213*\sy) -- (93.750000*\sx, 321.482213*\sy) -- cycle;
\fill[R150G150B255] (72.500000*\sx, 321.482213*\sy) -- (72.500000*\sx, 346.371941*\sy) -- (93.750000*\sx, 346.371941*\sy) -- (93.750000*\sx, 321.482213*\sy) -- (72.500000*\sx, 321.482213*\sy) -- cycle;
\draw[black] (72.500000*\sx, 321.482213*\sy) -- (72.500000*\sx, 346.371941*\sy) -- (93.750000*\sx, 346.371941*\sy) -- (93.750000*\sx, 321.482213*\sy) -- (72.500000*\sx, 321.482213*\sy) -- cycle;
\fill[R150G150B255] (72.500000*\sx, 253.482213*\sy) -- (72.500000*\sx, 321.482213*\sy) -- (115.000000*\sx, 321.482213*\sy) -- (115.000000*\sx, 253.482213*\sy) -- (72.500000*\sx, 253.482213*\sy) -- cycle;
\draw[black] (72.500000*\sx, 253.482213*\sy) -- (72.500000*\sx, 321.482213*\sy) -- (115.000000*\sx, 321.482213*\sy) -- (115.000000*\sx, 253.482213*\sy) -- (72.500000*\sx, 253.482213*\sy) -- cycle;
\fill[R150G150B255] (56.562500*\sx, 301.565474*\sy) -- (56.562500*\sx, 325.607105*\sy) -- (72.500000*\sx, 325.607105*\sy) -- (72.500000*\sx, 301.565474*\sy) -- (56.562500*\sx, 301.565474*\sy) -- cycle;
\draw[black] (56.562500*\sx, 301.565474*\sy) -- (56.562500*\sx, 325.607105*\sy) -- (72.500000*\sx, 325.607105*\sy) -- (72.500000*\sx, 301.565474*\sy) -- (56.562500*\sx, 301.565474*\sy) -- cycle;
\fill[R150G150B255] (51.250000*\sx, 253.482213*\sy) -- (51.250000*\sx, 301.565474*\sy) -- (72.500000*\sx, 301.565474*\sy) -- (72.500000*\sx, 253.482213*\sy) -- (51.250000*\sx, 253.482213*\sy) -- cycle;
\draw[black] (51.250000*\sx, 253.482213*\sy) -- (51.250000*\sx, 301.565474*\sy) -- (72.500000*\sx, 301.565474*\sy) -- (72.500000*\sx, 253.482213*\sy) -- (51.250000*\sx, 253.482213*\sy) -- cycle;
\fill[R150G150B255] (115.000000*\sx, 157.315691*\sy) -- (115.000000*\sx, 253.482213*\sy) -- (200.000000*\sx, 253.482213*\sy) -- (200.000000*\sx, 157.315691*\sy) -- (115.000000*\sx, 157.315691*\sy) -- cycle;
\draw[black] (115.000000*\sx, 157.315691*\sy) -- (115.000000*\sx, 253.482213*\sy) -- (200.000000*\sx, 253.482213*\sy) -- (200.000000*\sx, 157.315691*\sy) -- (115.000000*\sx, 157.315691*\sy) -- cycle;
\fill[R150G150B255] (157.500000*\sx, 109.232430*\sy) -- (157.500000*\sx, 157.315691*\sy) -- (200.000000*\sx, 157.315691*\sy) -- (200.000000*\sx, 109.232430*\sy) -- (157.500000*\sx, 109.232430*\sy) -- cycle;
\draw[black] (157.500000*\sx, 109.232430*\sy) -- (157.500000*\sx, 157.315691*\sy) -- (200.000000*\sx, 157.315691*\sy) -- (200.000000*\sx, 109.232430*\sy) -- (157.500000*\sx, 109.232430*\sy) -- cycle;
\fill[R150G150B255] (115.000000*\sx, 122.116301*\sy) -- (115.000000*\sx, 157.315691*\sy) -- (157.500000*\sx, 157.315691*\sy) -- (157.500000*\sx, 122.116301*\sy) -- (115.000000*\sx, 122.116301*\sy) -- cycle;
\draw[black] (115.000000*\sx, 122.116301*\sy) -- (115.000000*\sx, 157.315691*\sy) -- (157.500000*\sx, 157.315691*\sy) -- (157.500000*\sx, 122.116301*\sy) -- (115.000000*\sx, 122.116301*\sy) -- cycle;
\fill[R150G150B255] (72.500000*\sx, 185.482213*\sy) -- (72.500000*\sx, 253.482213*\sy) -- (115.000000*\sx, 253.482213*\sy) -- (115.000000*\sx, 185.482213*\sy) -- (72.500000*\sx, 185.482213*\sy) -- cycle;
\draw[black] (72.500000*\sx, 185.482213*\sy) -- (72.500000*\sx, 253.482213*\sy) -- (115.000000*\sx, 253.482213*\sy) -- (115.000000*\sx, 185.482213*\sy) -- (72.500000*\sx, 185.482213*\sy) -- cycle;
\fill[R150G150B255] (93.750000*\sx, 151.482213*\sy) -- (93.750000*\sx, 185.482213*\sy) -- (115.000000*\sx, 185.482213*\sy) -- (115.000000*\sx, 151.482213*\sy) -- (93.750000*\sx, 151.482213*\sy) -- cycle;
\draw[black] (93.750000*\sx, 151.482213*\sy) -- (93.750000*\sx, 185.482213*\sy) -- (115.000000*\sx, 185.482213*\sy) -- (115.000000*\sx, 151.482213*\sy) -- (93.750000*\sx, 151.482213*\sy) -- cycle;
\fill[R150G150B255] (72.500000*\sx, 160.592486*\sy) -- (72.500000*\sx, 185.482213*\sy) -- (93.750000*\sx, 185.482213*\sy) -- (93.750000*\sx, 160.592486*\sy) -- (72.500000*\sx, 160.592486*\sy) -- cycle;
\draw[black] (72.500000*\sx, 160.592486*\sy) -- (72.500000*\sx, 185.482213*\sy) -- (93.750000*\sx, 185.482213*\sy) -- (93.750000*\sx, 160.592486*\sy) -- (72.500000*\sx, 160.592486*\sy) -- cycle;
\fill[R150G150B255] (51.250000*\sx, 205.398952*\sy) -- (51.250000*\sx, 253.482213*\sy) -- (72.500000*\sx, 253.482213*\sy) -- (72.500000*\sx, 205.398952*\sy) -- (51.250000*\sx, 205.398952*\sy) -- cycle;
\draw[black] (51.250000*\sx, 205.398952*\sy) -- (51.250000*\sx, 253.482213*\sy) -- (72.500000*\sx, 253.482213*\sy) -- (72.500000*\sx, 205.398952*\sy) -- (51.250000*\sx, 205.398952*\sy) -- cycle;
\fill[R150G150B255] (56.562500*\sx, 181.357322*\sy) -- (56.562500*\sx, 205.398952*\sy) -- (72.500000*\sx, 205.398952*\sy) -- (72.500000*\sx, 181.357322*\sy) -- (56.562500*\sx, 181.357322*\sy) -- cycle;
\draw[black] (56.562500*\sx, 181.357322*\sy) -- (56.562500*\sx, 205.398952*\sy) -- (72.500000*\sx, 205.398952*\sy) -- (72.500000*\sx, 181.357322*\sy) -- (56.562500*\sx, 181.357322*\sy) -- cycle;
\definecolor{gray230}{RGB}{230,230,230};
\definecolor{gray128}{RGB}{128,128,128};
\fill[gray128] (400.000000*\sx, 8.721922*\sy) -- (400.000000*\sx, 4.927189*\sy) -- (0.000000*\sx, 4.927189*\sy) -- (0.000000*\sx, 8.721922*\sy) -- cycle;
\fill[gray128] (6.500000*\sx, 506.464426*\sy) -- (3.500000*\sx, 506.464426*\sy) -- (3.500000*\sx, 0.500000*\sy) -- (6.500000*\sx, 0.500000*\sy) -- cycle;
\draw[gray128] (30.000000*\sx,3.029823*\sy) -- (30.000000*\sx, 10.619289*\sy);
\definecolor{gray64}{RGB}{64,64,64};
\node[align=center,text=gray64,font=\tiny] at (30.000000*\sx, 19.473666*\sy) { 0 };
\draw[gray128] (115.000000*\sx,3.029823*\sy) -- (115.000000*\sx, 10.619289*\sy);
\node[align=center,text=gray64,font=\tiny] at (115.000000*\sx, 19.473666*\sy) { 0.5 };
\draw[gray128] (200.000000*\sx,3.029823*\sy) -- (200.000000*\sx, 10.619289*\sy);
\node[align=center,text=gray64,font=\tiny] at (200.000000*\sx, 19.473666*\sy) { 1 };
\draw[gray128] (285.000000*\sx,3.029823*\sy) -- (285.000000*\sx, 10.619289*\sy);
\node[align=center,text=gray64,font=\tiny] at (285.000000*\sx, 19.473666*\sy) { 1.5 };
\draw[gray128] (370.000000*\sx,3.029823*\sy) -- (370.000000*\sx, 10.619289*\sy);
\node[align=center,text=gray64,font=\tiny] at (370.000000*\sx, 19.473666*\sy) { 2 };
\draw[gray128] (455.000000*\sx,3.029823*\sy) -- (455.000000*\sx, 10.619289*\sy);
\node[align=center,text=gray64,font=\tiny] at (455.000000*\sx, 19.473666*\sy) { 2.5 };
\draw[gray128] (2.000000*\sx,-18.517787*\sy) -- (8.000000*\sx, -18.517787*\sy);
\node[align=center,text=gray64,font=\tiny] at (5.000000*\sx, -18.517787*\sy) { -1 };
\draw[gray128] (2.000000*\sx,117.482213*\sy) -- (8.000000*\sx, 117.482213*\sy);
\node[align=center,text=gray64,font=\tiny] at (5.000000*\sx, 117.482213*\sy) { -0.5 };
\draw[gray128] (2.000000*\sx,253.482213*\sy) -- (8.000000*\sx, 253.482213*\sy);
\node[align=center,text=gray64,font=\tiny] at (5.000000*\sx, 253.482213*\sy) { 0 };
\draw[gray128] (2.000000*\sx,389.482213*\sy) -- (8.000000*\sx, 389.482213*\sy);
\node[align=center,text=gray64,font=\tiny] at (5.000000*\sx, 389.482213*\sy) { 0.5 };
\draw[gray128] (2.000000*\sx,525.482213*\sy) -- (8.000000*\sx, 525.482213*\sy);
\node[align=center,text=gray64,font=\tiny] at (5.000000*\sx, 525.482213*\sy) { 1 };
\node[align=center,text=black,font=\tiny] at (200.000000*\sx, 0.500000*\sy) { x };
\node[align=center,text=black,font=\tiny] at (0.000000*\sx, 253.482213*\sy) { y };
\end{tikzpicture}

%% file: graphs.tex
The constraint hypergraph of a constraint satisfaction problem is a
hypergraph in which the vertices correspond to the variables, and the
hyperedges correspond to the constraints. Hypergraph representations
can be integrated with various constraint-solving algorithms, such as
backtracking, constraint propagation, and local search. The hypergraph
structure guides the search for a solution. For our needs, an
interesting idea is the detection of connected components.
Indeed, the partition of the constraint hypergraph into connected components
correspond to a partition of the problem into statistically
independent sub problems.
This means that the sampling for each component can be decorrelated from the
sampling of the others.

For instance, suppose we want to generate diagonally dominant
matrices.  A matrix is said to be diagonally dominant when, in every
row, the magnitude of the diagonal element is greater than or equal to
the sum of the magnitudes of the remaining elements in that
row. Formally, for each row \(i\),
\[
|a_{ii}| \ge \sum_{j \ne i} |a_{ij}|.
\]

\begin{figure}[h]
    \centering
\begin{tikzpicture}
\begin{scope}[every node/.style={circle,thick,draw}]
\node (v14) at (-0.223,3.920) {$v_{14}$};
\node (v15) at (0.923,5.660) {$v_{15}$};
\node (v13) at (1.943,2.580) {$v_{13}$};
\node (v12) at (0.823,0.940) {$v_{12}$};

\node (v8)  at (2.963,3.920) {$v_8$};
\node (v9)  at (4.163,5.660) {$v_9$};
\node (v11) at (5.183,2.580) {$v_{11}$};
\node (v10) at (4.063,0.940) {$v_{10}$};

\node (v6)  at (6.203,3.920) {$v_6$};
\node (v7)  at (7.403,5.660) {$v_7$};
\node (v5)  at (8.403,2.580) {$v_5$};
\node (v4)  at (7.303,0.940) {$v_4$};

\node (v2)  at (9.403,3.920) {$v_2$};
\node (v3)  at (10.603,5.660) {$v_3$};
\node (v1)  at (11.603,2.580) {$v_1$};
\node (v0)  at (10.503,0.940) {$v_0$};
\end{scope}

\begin{scope}[>={Stealth[black]},
              every node/.style={fill=white,circle},
              every edge/.style={draw=red,thick}]
  \path [-] (v14) edge node {$c_0$} (v15);
  \path [-] (v13) edge node {$c_0$} (v14);
  \path [-] (v13) edge node {$c_0$} (v15);
  \path [-] (v12) edge node {$c_0$} (v14);
  \path [-] (v12) edge node {$c_0$} (v15);
  \path [-] (v12) edge node {$c_0$} (v13);

  \path [-] (v8) edge node {$c_1$} (v9);
  \path [-] (v11) edge node {$c_1$} (v8);
  \path [-] (v11) edge node {$c_1$} (v9);
  \path [-] (v10) edge node {$c_1$} (v8);
  \path [-] (v10) edge node {$c_1$} (v9);
  \path [-] (v10) edge node {$c_1$} (v11);

  \path [-] (v6) edge node {$c_2$} (v7);
  \path [-] (v5) edge node {$c_2$} (v6);
  \path [-] (v5) edge node {$c_2$} (v7);
  \path [-] (v4) edge node {$c_2$} (v6);
  \path [-] (v4) edge node {$c_2$} (v7);
  \path [-] (v4) edge node {$c_2$} (v5);

  \path [-] (v2) edge node {$c_3$} (v3);
  \path [-] (v1) edge node {$c_3$} (v2);
  \path [-] (v1) edge node {$c_3$} (v3);
  \path [-] (v0) edge node {$c_3$} (v2);
  \path [-] (v0) edge node {$c_3$} (v3);
  \path [-] (v0) edge node {$c_3$} (v1);

\end{scope}

\end{tikzpicture}
\caption{Constraint graph for DDMs of size 4\label{fig:ddm}}
\end{figure}
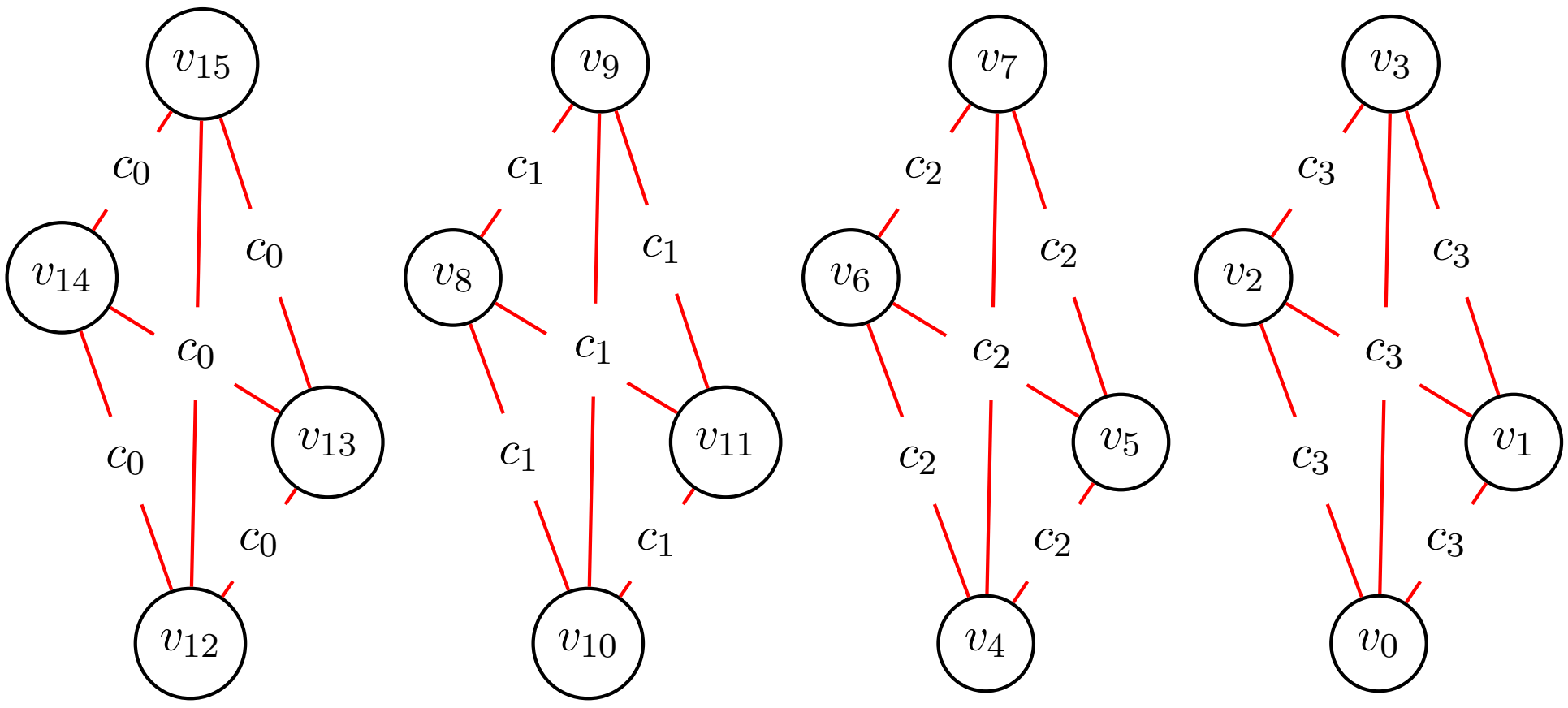

If we want to generate matrices of size 4, this means we have to solve
a CSP with 16 variables, that is four per row, and 4 constraints, that
is one per row. The diagonal dominance condition is independent for
each row. This is highlighted by the constraint graph of such a CSP
in \ref{fig:ddm}. This means that a randomly generated matrix may have
some rows that already satisfy it. In such cases, we can keep the rows
that are diagonally dominant and only regenerate the rows that violate
the condition, instead of discarding the entire matrix. This allows us
to preserve valid structure and reduces unnecessary computation.

In other words, if the constraint graph consists of two (or more)
connected components, it is then possible to generate solutions
independently in the different components, and then combine the
results.  Intuitively, leveraging this independence should speed up
the random generation process.

We also give a heuristic argument. The acceptance rate of a set of
constraints $\mathcal{C}$ is the probability that every constraint in
$\mathcal{C}$ accepts a random sample.  We note denote by $P_{\mathcal
C}$ this probability.  The expected number of rejections before a
sample satisfies all the constraints in~$\mathcal C$ is
thus~$\frac{1}{P_{\mathcal C}}$.  When there are two (or more)
disjoint components in the graph, the set $\mathcal{C}$ can be split
into two independent subsets, $\mathcal C_1$ and $\mathcal C_2$, with
respective supports $\mathcal V_1$ and $\mathcal V_2$ and we
have~$P_{\mathcal C} = P_{\mathcal C_1} P_{\mathcal C_2}$.  Leveraging
the independence of the variables in~$\mathcal V_1$ and~$\mathcal V_2$
by performing the rejection sampling independently for the two
components, yields a expected number of rejections of the order of
\begin{equation*}
  \frac{1}{P_{\mathcal C_1}} +
  \frac{1}{P_{\mathcal C_2}}
  \qquad\text{instead of}\qquad
  \frac{1}{P_{\mathcal C_1} P_{\mathcal C_2}}
\end{equation*}
which is largely smaller when the constraints~${\mathcal C_i}$ are hard to
satisfy, \textit{i.e.} when the~$P_{\mathcal C_i}$ are small.
Of course, a more precise analysis would require to take into account the cost
of sampling a leaf in the Huffman tree and the cost of generating a uniform
sample inside an abstract element.
However, we can already foresee that splitting the Huffman tree into the two
trees for the two connected components, and reducing the number of variables
will only moderately affect the performance in the algorithm.
We can conclude with confidence that leveraging the independence of the
components, will be beneficial.

In practice, at initialization, the CSP is divided into (disjoint) connected
components, and a sampling cache is associated with each one. This cache stores
the partial samples generated for each component, allowing the algorithm to
reuse previously computed samples and avoid redundant computations.

We benefit from this idea by using the cache to identify and
prioritize the solver's efforts on components that exhibit high
rejection rates. Empty caches indicate difficulty in finding valid
samples, encouraging the solver to select a variable to split in
these components. By adapting to the performance of individual
components, the solver converges toward a locally optimal
configuration.

\subsection{The \emph{bridge-first} split}
The iterative process of splitting and filtering gradually ensures
that certain constraints are locally satisfied, allowing them to be
removed from the graph. Whenever a constraint is removed, the solver
checks whether its removal disconnects the graph. When this is the case, the
graph is split into connected components, enabling their independent
handling.

Constraints become redundant when their validity is guaranteed by
current domain assignments. Our implementation recomputes connected
components whenever a constraint is removed by the solver so as to be
able to exploit the independence of the components.

To achieve this, we develop a propagation scheme and an exploration
heuristic that help reduce the connectivity of the constraint
graph. For each edge, we maintain information on whether it is a
bridge or not. A bridge is an edge whose removal increases the number
of connected components in the graph. We identify bridges using
Tarjan's algorithm \cite{Ta74}. Tracking which edges are bridges is
useful for guiding exploration, since variables that are part of a
bridge are particularly interesting for splitting. When a constraint
is removed, if it was a bridge, the connected components are
recomputed. Otherwise, the removal is checked to determine whether it
introduced new bridges, without recalculating the connected
components. If a bridge is detected, we attempt to eliminate it, as
its removal would disconnect the graph and thus improve the sampler's
performance.

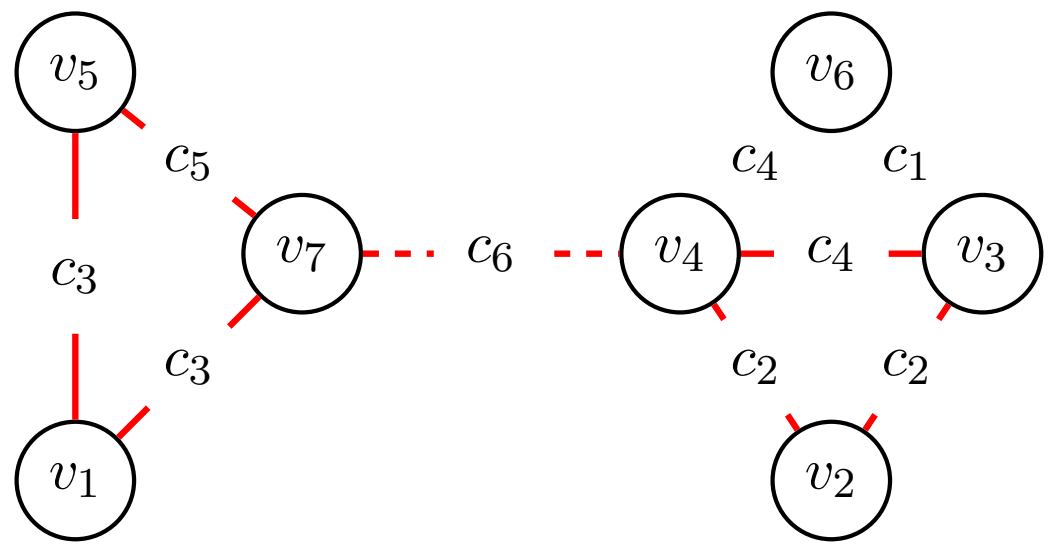
\begin{figure}[t]
    \centering
    \begin{tikzpicture}
      \begin{scope}[every node/.style={circle,thick,draw}]
        \node (V1) at (-3,0.5) {$v_1$};
        \node (V2) at (2,0.5)  {$v_2$};
        \node (V3) at (3,2)  {$v_3$};
        \node (V4) at (1,2)  {$v_4$};
        \node (V5) at (-3,3.2) {$v_5$};
        \node (V6) at (2,3.2)  {$v_6$};
        \node (V7) at (-1.5,2) {$v_7$};
      \end{scope}
      
      \begin{scope}[>={Stealth[black]},
          every node/.style={fill=white,circle},
          every edge/.style={draw=red,very thick}]
        \path [-] (V4) edge node {$c_4$} (V6);
        \path [-] (V3) edge node {$c_1$} (V6);
        \path [-] (V2) edge node {$c_2$} (V3);
        \path [-] (V2) edge node {$c_2$} (V4);
        \path [-] (V1) edge node {$c_3$} (V7);
        \path [-] (V1) edge node {$c_3$} (V5);
        \path [-] (V4) edge node {$c_4$} (V3);
        \path [-] (V7) edge node {$c_5$} (V5);
        \path [-,dashed] (V7) edge node {$c_6$} (V4);
      \end{scope}
    \end{tikzpicture}
    \caption{Deletion of the constraint $c_6$ leads to two independent CSPs.}
\end{figure}

At initialization, we compute the bridges information for each
connected component and use it to populate a priority queue of
variables. During exploration, the solver gives priority to variables
in this queue: with some probability, it selects the next prioritized
variable and splits the abstract domain along it; otherwise, it uses
the usual splitting heuristic. This biases exploration towards
variables whose splitting that may help reduce graph connectivity,
while retaining the possibility of exploring other variables. The
bridge information is also recomputed when a component is disconnected
into smaller components.

\subsection{Sharing constraint graphs across subproblems}

The exploration process may generate many subproblems from the same
initial CSP. Although each subproblem has its own abstract domain, its
constraint graph often remains identical to that of its parent, at
least until propagation removes some constraints. Constructing and
maintaining a separate copy of the graph for every subproblem would
therefore introduce unnecessary work and memory overhead.

Our implementation avoids this duplication by sharing graph states
across branches of the exploration tree. When a component is split,
the resulting subproblems retain the same constraint graph as their
parent; only their abstract domains and splitting information
differ. When propagation removes a constraint, a new graph state is
obtained by removing the corresponding edge. Moreover, this operation
is memoized: if another branch removes the same constraint from the
same graph state, the previously computed result can be reused. We
apply the same idea to graph analyses, memoizing the computation of
connected components and bridge detection.

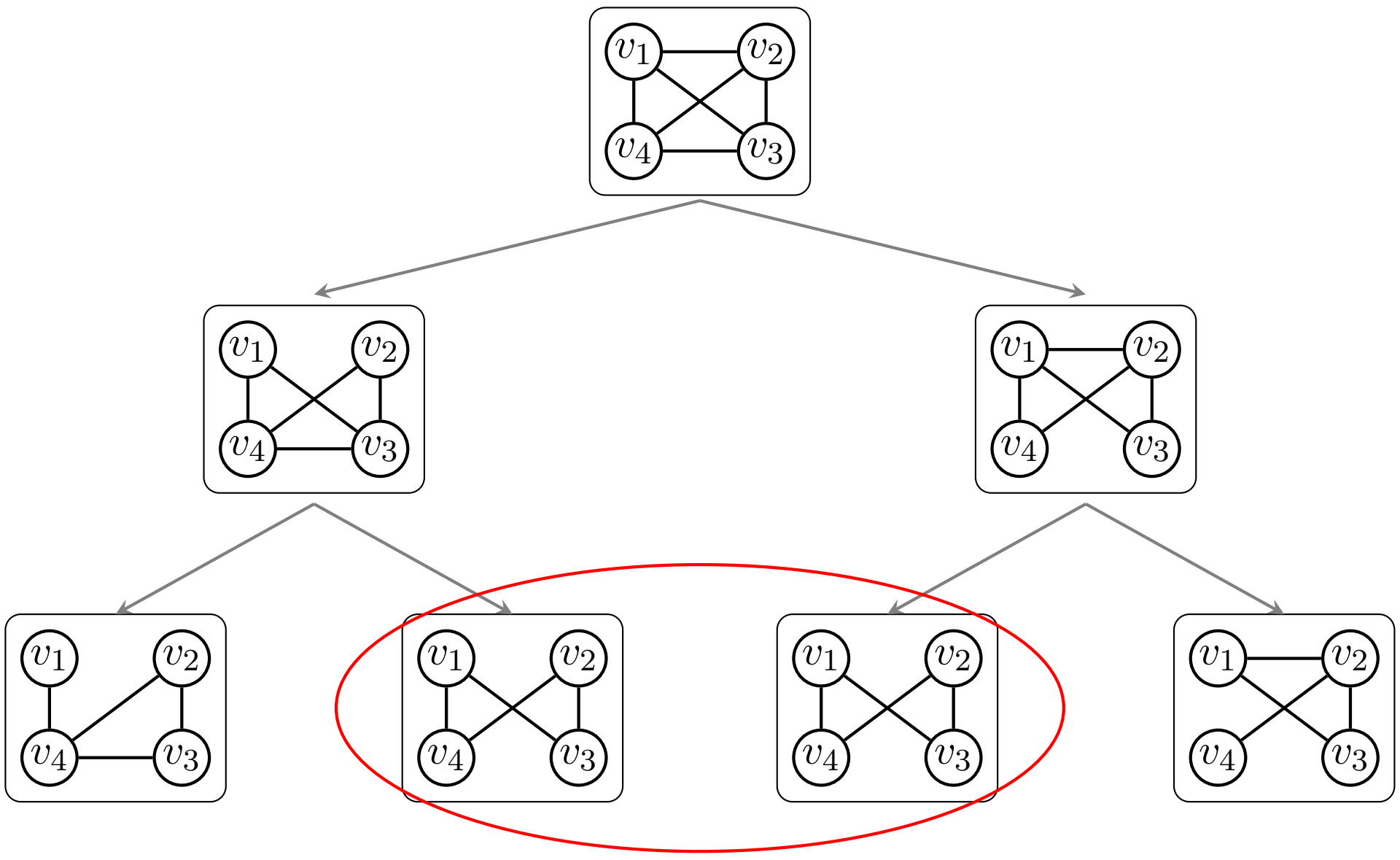
\begin{figure}[t]
    \centering
    \begin{tikzpicture}[
        vertex/.style={
            circle,
            draw,
            thick,
            minimum size=5mm,
            inner sep=0pt
        },
        edge/.style={thick},
        tree edge/.style={
            ->,
            >=stealth,
            thick,
            gray
        },
        graph box/.style={
            draw,
            rounded corners,
            inner sep=5pt
        }
    ]

    \begin{scope}[shift={(0,0)}]
        \node[vertex] (r1) at (-0.6,0.45) {$v_1$};
        \node[vertex] (r2) at ( 0.6,0.45) {$v_2$};
        \node[vertex] (r3) at ( 0.6,-0.45) {$v_3$};
        \node[vertex] (r4) at (-0.6,-0.45) {$v_4$};

        \draw[edge] (r1)--(r2);
        \draw[edge] (r1)--(r3);
        \draw[edge] (r1)--(r4);
        \draw[edge] (r2)--(r3);
        \draw[edge] (r2)--(r4);
        \draw[edge] (r3)--(r4);

        \draw[graph box] (-1,-0.85) rectangle (1,0.85);
    \end{scope}

    \begin{scope}[shift={(-3.5,-2.7)}]
        \node[vertex] (a1) at (-0.6,0.45) {$v_1$};
        \node[vertex] (a2) at ( 0.6,0.45) {$v_2$};
        \node[vertex] (a3) at ( 0.6,-0.45) {$v_3$};
        \node[vertex] (a4) at (-0.6,-0.45) {$v_4$};

        \draw[edge] (a1)--(a3);
        \draw[edge] (a1)--(a4);
        \draw[edge] (a2)--(a3);
        \draw[edge] (a2)--(a4);
        \draw[edge] (a3)--(a4);

        \draw[graph box] (-1,-0.85) rectangle (1,0.85);
    \end{scope}

    \begin{scope}[shift={(3.5,-2.7)}]
        \node[vertex] (b1) at (-0.6,0.45) {$v_1$};
        \node[vertex] (b2) at ( 0.6,0.45) {$v_2$};
        \node[vertex] (b3) at ( 0.6,-0.45) {$v_3$};
        \node[vertex] (b4) at (-0.6,-0.45) {$v_4$};

        \draw[edge] (b1)--(b2);
        \draw[edge] (b1)--(b3);
        \draw[edge] (b1)--(b4);
        \draw[edge] (b2)--(b3);
        \draw[edge] (b2)--(b4);

        \draw[graph box] (-1,-0.85) rectangle (1,0.85);
    \end{scope}

    \begin{scope}[shift={(-5.3,-5.5)}]
        \node[vertex] (c1) at (-0.6,0.45) {$v_1$};
        \node[vertex] (c2) at ( 0.6,0.45) {$v_2$};
        \node[vertex] (c3) at ( 0.6,-0.45) {$v_3$};
        \node[vertex] (c4) at (-0.6,-0.45) {$v_4$};

        \draw[edge] (c1)--(c4);
        \draw[edge] (c2)--(c3);
        \draw[edge] (c2)--(c4);
        \draw[edge] (c3)--(c4);

        \draw[graph box] (-1,-0.85) rectangle (1,0.85);
    \end{scope}

    \begin{scope}[shift={(-1.7,-5.5)}]
        \node[vertex] (d1) at (-0.6,0.45) {$v_1$};
        \node[vertex] (d2) at ( 0.6,0.45) {$v_2$};
        \node[vertex] (d3) at ( 0.6,-0.45) {$v_3$};
        \node[vertex] (d4) at (-0.6,-0.45) {$v_4$};

        \draw[edge] (d1)--(d3);
        \draw[edge] (d1)--(d4);
        \draw[edge] (d2)--(d3);
        \draw[edge] (d2)--(d4);

        \draw[graph box] (-1,-0.85) rectangle (1,0.85);
    \end{scope}

    \begin{scope}[shift={(1.7,-5.5)}]
        \node[vertex] (e1) at (-0.6,0.45) {$v_1$};
        \node[vertex] (e2) at ( 0.6,0.45) {$v_2$};
        \node[vertex] (e3) at ( 0.6,-0.45) {$v_3$};
        \node[vertex] (e4) at (-0.6,-0.45) {$v_4$};

        \draw[edge] (e1)--(e4);
        \draw[edge] (e2)--(e3);
        \draw[edge] (e1)--(e3);
        \draw[edge] (e2)--(e4);

        \draw[graph box] (-1,-0.85) rectangle (1,0.85);
    \end{scope}

    \begin{scope}[shift={(5.3,-5.5)}]
        \node[vertex] (f1) at (-0.6,0.45) {$v_1$};
        \node[vertex] (f2) at ( 0.6,0.45) {$v_2$};
        \node[vertex] (f3) at ( 0.6,-0.45) {$v_3$};
        \node[vertex] (f4) at (-0.6,-0.45) {$v_4$};

        \draw[edge] (f1)--(f2);
        \draw[edge] (f1)--(f3);
        \draw[edge] (f2)--(f3);
        \draw[edge] (f2)--(f4);

        \draw[graph box] (-1,-0.85) rectangle (1,0.85);
    \end{scope}

    \draw[tree edge] (0,-0.9) -- (-3.5,-1.75);
    \draw[tree edge] (0,-0.9) -- (3.5,-1.75);

    \draw[tree edge] (-3.5,-3.65) -- (-5.3,-4.65);
    \draw[tree edge] (-3.5,-3.65) -- (-1.7,-4.65);

    \draw[tree edge] (3.5,-3.65) -- (1.7,-4.65);
    \draw[tree edge] (3.5,-3.65) -- (5.3,-4.65);

    \draw[red, thick] (0,-5.5) ellipse (3.3cm and 1.3cm);
\end{tikzpicture}

    \caption{Evolution of constraint graphs along an exploration
    tree. Each child graph is obtained from its parent after at least
    one propagation/exploration cycle, during which one or more
    constraints may be removed. Different branches may lead to identical
    graph states.}  \label{fig:graph-sharing}
\end{figure}

Figure~\ref{fig:graph-sharing} illustrates this mechanism. The root
represents the initial constraint graph. Each child is obtained by
removing one or more edges from its parent, and therefore contains a
subset of its parent's edges. Different branches may reuse graph
states and the results of graph operations whenever they encounter the
same configuration.

%% file: incremental.tex
One key idea for improving the construction of generators is to reuse
the solving state of a generator when moving from an instance of size
$n$ to a larger instance of size $n+1$. This situation commonly occurs
when generating structured data, such as lists, trees, or matrices,
for which the generator is invoked at different sizes. In the current
approach, each problem instance may require the construction and
refinement of a new abstract cover, even when the larger instance
shares most of its structure and constraints with the smaller
one. Reusing the knowledge acquired at dimension $n$ could therefore
significantly reduce the cost of constructing a generator at dimension
$n+1$.

\subsection{Problem Formulation}

Let
\[
C_n = (V_n, D_n, \mathcal{C}_n)
\]
denote a CSP of size $n$, where $V_n$ is its set of variables, $D_n$ their
domains, and $\mathcal{C}_n$ its set of constraints. We consider a larger
CSP
\[
C_{n+1} = (V_{n+1}, D_{n+1}, \mathcal{C}_{n+1}),
\]
obtained by adding one or more variables and the constraints required to
integrate them into the problem.

In the simplest case, we have
\[
V_{n+1} = V_n \cup \{x_{n+1}\}
\]
and
\[
\mathcal{C}_{n+1} = \mathcal{C}_n \cup \Delta\mathcal{C},
\]
where $\Delta\mathcal{C}$ contains the new constraints introduced by the
additional variable. The objective is to construct a generator for
$C_{n+1}$ by extending the solving state already computed for $C_n$,
rather than solving the larger problem from scratch.

\subsection{Reusing the Abstract Cover}

Recall that our solving procedure maintains an abstract cover of the
solution space, composed of inner elements $I_n$ and outer elements
$O_n$. These elements represent knowledge acquired during the refinement
process: inner elements are known to contain only solutions, while outer
elements represent regions that may still contain solutions.

The first step is to lift these elements into the larger search
space. For an abstract element $B$ over the variables $V_n$, its
extension can be defined as
\[
B' = B \times D(x_{n+1}),
\]
where $D(x_{n+1})$ is the domain of the newly introduced variable.

This operation provides an initial cover of the larger search space using
the information already computed at dimension $n$. However, the lifted
elements cannot automatically be considered solutions of $C_{n+1}$:
the new constraints may invalidate some or all of their assignments.
Consequently, the extended cover must be filtered against
$\Delta\mathcal{C}$, and the resulting outer elements must be refined when
necessary.

The advantage is that the solver does not need to rediscover the structure
of the original search space. It can start from the previously acquired
representation and focus its efforts on the additional dimensions and
constraints.

\subsection{Detecting and Reusing Previously Processed Constraints}

The reuse mechanism can be implemented directly at the level of the
constraint-solving engine by maintaining a cache of previously
constructed CSPs together with their solving states. When a new CSP is
constructed, the cache is searched for a compatible smaller instance
whose solving state can be reused.

More precisely, two CSPs are considered compatible when all the
variables and constraints of the smaller CSP are also present in the
larger one. Let $C_s$ denote a cached CSP and $C_n$ the newly constructed
CSP. We say that $C_s$ is included in $C_n$ if
$$
V_s \subseteq V_n
\qquad\text{and}\qquad
\mathcal{C}_s \subseteq \mathcal{C}_n.
$$
The additional components are then given by
$$\Delta V = V_n \setminus V_s
\qquad\text{and}\qquad
\Delta\mathcal{C} = \mathcal{C}_n \setminus \mathcal{C}_s.
$$

The cache stores, for each generator, a collection of previously
encountered CSPs and associated state accessors. Entries are maintained
in decreasing order of CSP size, measured here as the sum of the number
of variables and constraints. This ordering allows the search procedure
to select the largest cached CSP that is included in the new instance.
Reusing the largest compatible instance maximizes the amount of
previously computed solving information that can be retained.

Once such an instance has been found, its solving state is recovered
and extended incrementally with the additional variables and
constraints:

$$S_n = S_s \cup \Delta V \cup \Delta\mathcal{C}$$

where $S_s$ is the previously cached solving state. The sampler is
then reconstructed from this extended state, rather than initialized
from the complete CSP.

If no compatible cached instance exists, the solver falls back to
constructing a new solving state from the entire CSP. In either case,
the newly constructed CSP and its associated state accessor are added
to the cache, making them available for subsequent extensions.

An important aspect of this approach is that compatibility is
determined through structural inclusion of variables and constraints.
This assumes that constraints appearing in both instances have
identical semantics and that their previously computed solving
information remains valid when the additional variables and
constraints are introduced. In particular, a constraint whose meaning
depends on the complete variable collection must be represented
appropriately if its semantics change as the CSP grows.

This mechanism provides a simple form of incremental solving: instead
of rebuilding the solving state for every instance, the solver retains
the state of the largest compatible previously encountered instance
and processes only the structural additions. It is particularly useful
when successive CSPs share most of their variables and constraints,
as is often the case when generating increasingly large structured
objects.

\subsection{Preserving Uniformity}

One important consideration is that reusing a generator must not
compromise the uniformity of the generated solutions. In particular,
simply sampling a solution of $C_n$ and assigning a random value to the
new variable does not generally produce a uniform sample of
$C_{n+1}$.

Indeed, different solutions of $C_n$ may admit different numbers of
valid extensions to the larger problem. Assigning equal probability to
the old solutions would therefore introduce a bias whenever their
numbers of valid extensions differ.

The extended abstract cover must preserve the sampling properties of
the original method. In particular, if the resulting elements are
non-overlapping, their selection probabilities must account for the
volume of their valid regions. More generally, the probability
assigned to an element must be proportional to the number or measure
of solutions it represents. The usual sampling procedure can then be
applied to the refined cover, provided that the required volume
information and non-overlap conditions are maintained.

%% file: gegen.tex
We have implemented the techniques presented in this paper in a
prototype called \textit{GeGen}, which stands
for \textit{\textbf{Ge}nerator \textbf{Gen}erator}. PBT frameworks
like QCheck~\cite{qcheck}, which is the OCaml port of QuickCheck,
provide generators for atomic types (bool, ints, floats, \dots) and
combinators for composite types (pairs, tuples, \dots). In the
presence of recursive types, the traditional approach is to provide
the user with a generator of \textit{sized values}, that is, the user
has to provide a size, and the generator builds values with that
specific size. Most implementations lack robust support for testing
scenarios that require constraints over the generated
inputs. \textit{GeGen} bridges this gap by introducing variable
generators and constraints, enabling the easy integration of
constraint-solving capabilities in a PBT framework.

For instance, QCheck provides the function \lstinline{find_example}
whose signature is given below. Given a generator of values of type
$t$, it builds a generator of values of $t$ that satisfy a certain
predicate:

\begin{lstlisting}[caption={QCheck's find example function signature},
                   label={lst:find-example}]
find_example : ('a -> bool) -> 'a Gen.t -> 'a Gen.t
\end{lstlisting}

If a value satisfying the predicate \lstinline{f} is found (within a
certain number of tries), it is returned. \textit{GeGen} mimics this
approach by providing an API that users can manipulate
similarly. However, the inner mechanism differs notably, as we build,
solve, and sample from a CSP during the generation process.

\subsection{GeGen's Generators}

GeGen's generators differ fundamentally from traditional
QuickCheck-style generators as they produce \emph{symbolic variables}
rather than concrete values. Traditional property-based testing
frameworks generate fixed values like random integers or floats. In
contrast, GeGen's generators operate at a symbolic level, which makes
constraint composition and solving possible. For instance, instead of
directly generating a number, \textit{GeGen} creates a symbolic
variable that represents the number and associates it with a range of
possible values. This symbolic variable acts as a declaration on the
solver's side. Additionally, instead of applying a predicate to verify
whether a value satisfies the property being tested, GeGen's
predicates impose constraints on this variable, such as ``\textit{the
number must be even}'' or ``\textit{the number must be less than
another variable}'', which are collected to construct a CSP.

Finally, we apply the previously discussed solving and sampling
techniques to the constructed CSP. For example, to generate circles
within a given square in QCheck, we can use the following code:

\begin{lstlisting}[caption={Generating a circle within a square using QCheck},label={lst:gen-circle}]
find_example
  (pair (pair int int) int)
  (fun ((x, y), r) ->
    r > 0 && x >= r && x <= 100 - r && y >= r && y <= 100 - r && r >= 10)
\end{lstlisting}

The \lstinline{find_example} function generates random integer values
for \(x\), \(y\), and \(r\), then applies the given predicate to check
whether the generated circle satisfies the constraints. This process
repeats until a valid example is found or the search limit is
exhausted. Note that here, \lstinline{int} is an atomic generator that
produces random integers and \lstinline{pair} is a combinator that
constructs a generator of pairs from two base generators.

The interest of our approach is that the same code, when linked
against our library, will yield identical results but in a
significantly more efficient manner, using a fundamentally different
mechanism. Instead of directly sampling random values
for \(x\), \(y\), and \(r\), we construct a symbolic representation of
the problem. From the solver's perspective, a constraint satisfaction
problem (CSP) with three floating-point variables $v_1,v_2$ and $v_3$
is formulated. However, this approach abstracts certain details about
the algebraic structure of the type. Therefore, a reconstruction
function is designed alongside the CSP, so that once a sample $s$ is
drawn, it can be reassembled into a value of the correct type. In this
particular case, the function is
\[
f(s)=((s(v_1),s(v_2)),s(v_3)).
\]
It ensures that once the solver produces a valid assignment for the
symbolic variables, the corresponding concrete values are
reconstructed in a way that respects the intended algebraic
structure. Finally, the predicate, instead of returning a truth value,
actually builds the equivalent constraint system. In essence, this
allows us to decouple the problem-solving phase from the data
generation phase (at the API level) while ensuring that the generated
values maintain the correct type.

\subsection{GeGen's Language}

The constraint language used in \textit{GeGen} supports arithmetic and
Boolean expressions. It includes arithmetic operations (addition,
multiplication, etc.), Boolean logic (comparisons, conjunctions,
disjunctions, negations), and variables. We build upon those
generators for more complex types such as tuples and lists by
composing the different elements of a generator.

GeGen extends standard arithmetic and Boolean operators by overriding
them to facilitate the construction of constraints, making it possible
to define a CSP in a manner that closely resembles programming its
logical predicate counterpart. This intuitive approach aligns the
construction of constraints with familiar programming paradigms,
simplifying the transition from predicates to constraints.

For instance, consider a predicate that verifies whether a list is
sorted. This predicate can naturally be expressed in a functional
style as follows:

\begin{lstlisting}[caption={Functional definition of the sortedness predicate},
 label={lst:sorted}]
let rec is_sorted = function
| [] | [x] -> true
| x1 :: x2 :: rest -> x1 <= x2 && is_sorted (x2 :: rest)
\end{lstlisting}

This predicate can be applied to concrete lists of integers to
determine whether they are sorted. This is what is done when using
rejection sampling. Using GeGen, the predicate takes another meaning
as it operates on lists of variables to dynamically construct a
CSP. The overloaded operators automatically translate the predicate
logic into corresponding constraints. For example:

\begin{lstlisting}[caption={Encoding the sortedness predicate as a constraint},
                   label={lst:sorted-constraint}]
open GeGen
let rec is_sorted = function
| [] | [x] -> true_
| x1 :: x2 :: rest -> x1 <= x2 && is_sorted (x2 :: rest)
\end{lstlisting}

Here, the \lstinline{open GeGen} directive brings into scope the
operators \lstinline{<=} and \lstinline{&&} (among others), so that
instead of computing a Boolean value, the \lstinline{is_sorted}
function now builds a constraint. Besides that, the only syntactic
difference is the use of the symbolic constraint \lstinline{true_}
instead of the built-in OCaml boolean \lstinline{true}, which cannot
be avoided as OCaml does not permit the overriding of literals. This
approach preserves the logical structure and readability of the
predicate while generating a constraint representation that can be
used to solve CSPs.

\subsection{Gegen's Typical Use-Case}
GeGen's solver refines its internal state and search strategies over
time. Consequently, restarting the generation process from scratch at
every test run would be inefficient. To address this, GeGen provides
API operations to save and restore a generator's state. This allows
developers to persist the accumulated solving information between
test-suite executions and resume generation from an already-refined
state.

The following example illustrates a typical use-case. We define a
property stating that a list of integers must be sorted, and use our
tool to generate lists satisfying this property. The example follows a
familiar property-based testing (PBT) style: a generator describes the
input domain, a predicate specifies the property to satisfy, and
repeated calls to the resulting generator produce test cases.

The example considers lists of increasing sizes, from 2
to \texttt{max\_size}. For each size, a generator is obtained by
applying the same property to a list generator parameterized by that
size. It is then called repeatedly to produce test cases. From the
user's perspective, this is much like ordinary PBT: one defines a
generator and a property, then requests as many examples as needed.

However, the process differs substantially behind the scenes. Rather
than treating each generation request as an independent search, GeGen
relies on an incremental constraint solver. The predicate is
translated into constraints, and the solver maintains and refines an
internal state as it searches for satisfying inputs. As generation
proceeds, the solving information accumulated during previous calls
can be reused by later ones.

This gives rise to two levels of incrementality. First, each generator
improves over time: repeated requests for examples benefit from the
solver state refined during earlier requests. Second, generators for
different dimensions can reuse solving information across
dimensions. Since the generators are related, moving from one list
size to the next need not mean starting the solving process from
scratch. 

Finally, the accumulated state can be saved and restored through the
API. When the program starts, it attempts to load the previously saved
generator state. If no saved state is available, generation starts
with a fresh state. After the generation loop, the updated state is
saved for future executions. Thus, the incremental solving process can
continue across not only successive generation requests, but also
separate test-suite executions. The interface remains close to regular
PBT, while the underlying solver incrementally learns across calls,
across related generators, and across runs.

\begin{lstlisting}[caption={Generator State Management through the API},
                   label={lst:api}]
open Gegen

let rec is_sorted l =
  match l with
  | [] | [_] -> true_
  | x1 :: x2 :: rest -> x1 <= x2 && is_sorted (x2 :: rest)

let nb = 100

let count = 10000

let name = "sortedlist"

let cache_file = name ^ ".cache"

let max_size = 15

let _ =
  (* Load an existing generator, if available. *)
  if Sys.file_exists cache_file then (
    Format.printf "Loading generator %s...@." name ;
    load_generator ~name cache_file )
  else Format.printf "Creating generator %s...@." name ;
  let finder_sorted = find_example_cached ~count ~name in
  for i = 2 to max_size do
    let generator =
      finder_sorted Gen.(list_of_size (fun _ -> i) int) ~f:is_sorted
    in
    for _ = 0 to nb do
      generator () |> ignore
    done
  done ;
  (* Save the updated generator. *)
  Format.printf "Saving generator %s...@." name ;
  save_generator ~name cache_file
\end{lstlisting}

%% file: experiments.tex
Our implementation is written in OCaml in a functional style. We use
the default \textit{pseudo-random number generator} (PRNG) from the
OCaml standard library, an instance of the LXM \cite{St&al21} family
of PRNGs. \textit{GeGen} relies on the \textit{AbSolute}
solver \cite{Pe&al13} to handle constraints.  This solver provides
most of the necessary functionalities for uniform random sampling,
including volume measurement and efficient space paving with large
elements \cite{Zi&al18}. It guarantees a non-overlapping solution
cover, simplifying uniform distribution construction.  Our
implementation is open-source and available at: \\
{\small\url{https://osf.io/u4r5q/files/osfstorage?view_only=84af8ee65a6c495d98cb7b7bfad8c54b}}.

\subsection{Benchmark}
We have written generators using our framework for several realistic
examples: the \texttt{convex} problem defines the predicate for
star-convex polygons, \ie the vectors corresponding to two consecutive
edges of the polygons have a negative
cross-product. The \texttt{diagdom}, \texttt{idempotent},
\texttt{symmetric} and \texttt{orthogonal} problems, define the predicates for some well
known classes of matrices. The \texttt{distrib}, \texttt{itvcover}
and \texttt{sorted} represent respectively the predicate for valid
distribution of probability \ie, list of positive values that sum to
one, interval-covering list that is a list that contains a set of
intervals that cover a specific range without gaps and finally the set
of sorted lists in increasing order. We compare our approach against
QCheck's standard rejection sampler, as it serves as a baseline for
property-based testing. This comparison is relevant since we have
designed an API that allows users to write the same specifications
while seamlessly generating samples using our method instead.

Table~\ref{tab:times} presents the number of samples generated
by \textit{GeGen} and QCheck for various problem instances. Each row
corresponds to a specific problem. The first column lists the problem
names. The second column indicates the problem size parameters,
denoted as $|\mathcal{V}|,|\mathcal{C}|$. The remaining columns report
the number of generated samples for different time constraints (0.1,
1, and 2 seconds of generation time), with results shown separately
for \textit{GeGen} and QCheck. Best result for each are shown in bold
font. The problem sizes were selected to ensure that trends in sample
generation could be clearly observed while maintaining a fair
evaluation of both methods. The measurements were conducted on a
machine with a 12th Gen Intel(R) Core(TM) i5-1235U processor and 15
GiB of RAM.
\subsection{Results Analysis}
\floatplacement{table}{H}
\begin{table}
\centering
\begin{tabular}{|*{2}{c|}|*{3}{c|}|*{3}{c|}}
  \hline
 & & \multicolumn{3}{c|}{\textit{GeGen}} & \multicolumn{3}{c|}{QCheck}   \\ \hline
  problem    & $|\mathcal{V}|,|\mathcal{C}|$ & 0.1 &  1  & 2 &  0.1 & 1 & 2 \\ \hline
sortedlist4 & 4,3 & 
7076 & 292589 & 621378 & \textbf{17999} & \textbf{527764} & \textbf{1107631} \\ \hline
sortedlist8 & 8,7 & 
\textbf{410} & \textbf{32985} & \textbf{76783} & 9 & 1047 & 2191 \\ \hline
sortedlist12 & 12,11 & 
\textbf{28} & \textbf{4891} & \textbf{12421} & 0 & 0 & 0 \\ \hline
convex3 & 6,3 & 
\textbf{752} & \textbf{54829} & \textbf{114298} & 5 & 415 & 868 \\ \hline
convex4 & 8,4 & 
\textbf{203} & \textbf{9182} & \textbf{19997} & 0 & 0 & 2 \\ \hline
convex5 & 10,5 & 
\textbf{82} & \textbf{3526} & \textbf{7357} & 0 & 0 & 0 \\ \hline
diagdom3 & 9,4 & 
823 & \textbf{39282} & \textbf{83120} & \textbf{1052} & 25354 & 52677 \\ \hline
diagdom4 & 16,5 & 
\textbf{141} & \textbf{6454} & \textbf{13680} & 0 & 25 & 54 \\ \hline
diagdom5 & 25,6 & 
\textbf{20} & \textbf{984} & \textbf{2176} & 0 & 0 & 0 \\ \hline
itvcover3 & 6,5 & 
\textbf{5522} & \textbf{225252} & \textbf{477677} & 0 & 29 & 62 \\ \hline
itvcover4 & 8,4 & 
\textbf{4258} & \textbf{132492} & \textbf{279720} & 0 & 1 & 3 \\ \hline
itvcover5 & 10,9 & 
\textbf{2724} & \textbf{87048} & \textbf{174027} & 0 & 0 & 0 \\ \hline
symmetric2 & 4,1 & 
5617 & \textbf{428439} & \textbf{901930} & \textbf{7842} & 242921 & 493929 \\ \hline
symmetric3 & 9,3 & 
\textbf{7178} & \textbf{233311} & \textbf{490004} & 0 & 1 & 3 \\ \hline
symmetric4 & 16,6 & 
\textbf{2176} & \textbf{136995} & \textbf{280146} & 0 & 0 & 0 \\ \hline
orthogonal2 & 4,5 & 
\textbf{1457} & \textbf{59079} & \textbf{125073} & 0 & 0 & 0 \\ \hline
orthogonal3 & 9,8 & 
\textbf{132} & \textbf{5110} & \textbf{10693} & 0 & 0 & 0 \\ \hline
orthogonal4 & 16,12 & 
\textbf{3} & \textbf{596} & \textbf{1329} & 0 & 0 & 0 \\ \hline
idempotent2 & 4,10 & 
\textbf{1494} & \textbf{93988} & \textbf{205829} & 0 & 15 & 33 \\ \hline
idempotent3 & 9,29 & 
\textbf{35} & \textbf{3590} & \textbf{8295} & 0 & 0 & 0 \\ \hline
idempotent4 & 16,76 & 
0 & \textbf{41} & \textbf{158} & 0 & 0 & 0 \\ \hline
\end{tabular}
\caption{Number of samples per generation time\label{tab:times}}
\end{table}

The results indicate that \textit{GeGen} consistently outperforms
QCheck, generating significantly more samples across nearly all
problem types and sizes. While QCheck performs adequately for smaller
problem instances, its performance drops as problem complexity
increases, often failing to generate any samples for larger cases. In
contrast, \textit{GeGen} scales effectively, maintaining high sample
generation rates even for more complex problems. Also, the performance
of the rejection sampler remains stable over time while the ones of
the generator built using \textit{GeGen} improves as the process
progresses. This trend is illustrated in Figure~\ref{fig:evol}, which
depicts the time in seconds (y-axis) required to generate the $i$-th
sample (x-axis), for 10000 samples. \textit{GeGen}'s times (solid
line), and QCheck's (dashed line) are shown on a single problem,
sorted lists of size $n$, to highlight this feature of our generators,
although the same behaviour can be noticed on problems of
table~\ref{tab:times}.  Notable spikes in the curves correspond to
garbage collection cycles, which are particularly apparent for small
values of $n$.

{\centering
\begin{figure}[ht]
\begin{tabular}{ccc}
  \begin{subfigure}{0.33\textwidth}\centering
    \includegraphics[scale=0.25]{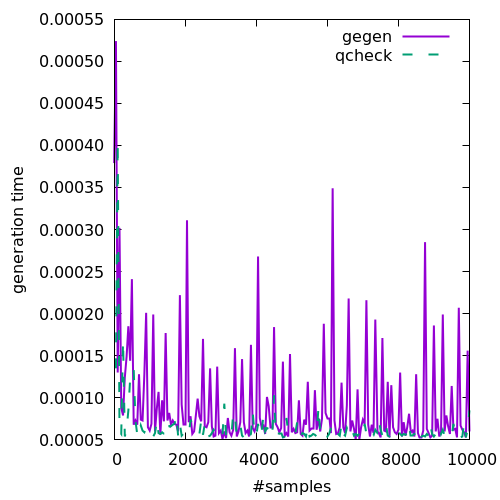}
    \caption{$n=3$\label{fig:evola}}
  \end{subfigure}&
  \begin{subfigure}{0.33\textwidth}\centering
    \includegraphics[scale=0.25]{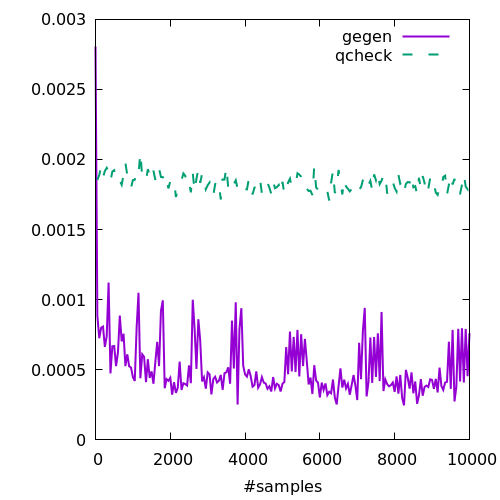}
    \caption{$n=6$\label{fig:evolb}}
   \end{subfigure}&
   \begin{subfigure}{0.30\textwidth}\centering
    \includegraphics[scale=0.25]{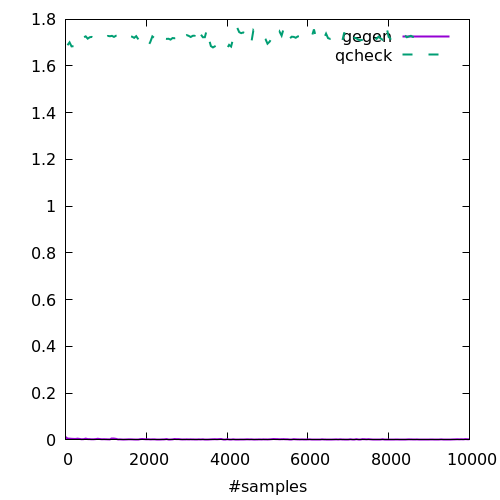}
    \caption{$n=9$\label{fig:evolc}}
   \end{subfigure}\\ \newline
   \begin{subfigure}{0.33\textwidth}\centering
    \includegraphics[scale=0.25]{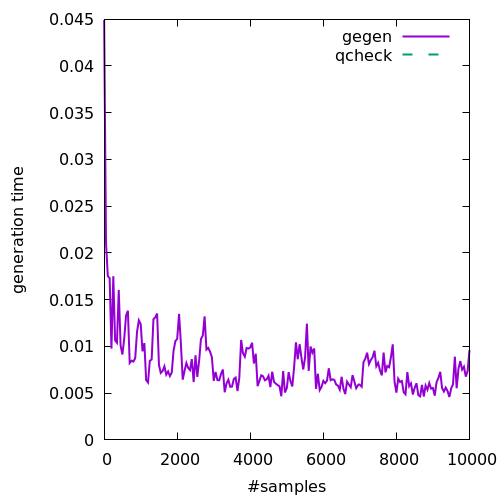}
     \caption{$n=12$\label{fig:evold}}
   \end{subfigure}&
  \begin{subfigure}{0.33\textwidth}\centering
    \includegraphics[scale=0.25]{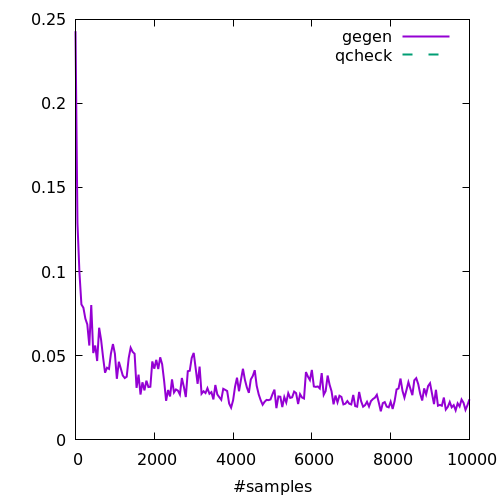}
    \caption{$n=15$\label{fig:evole}}
  \end{subfigure}&
  \begin{subfigure}{0.30\textwidth}\centering
    \includegraphics[scale=0.25]{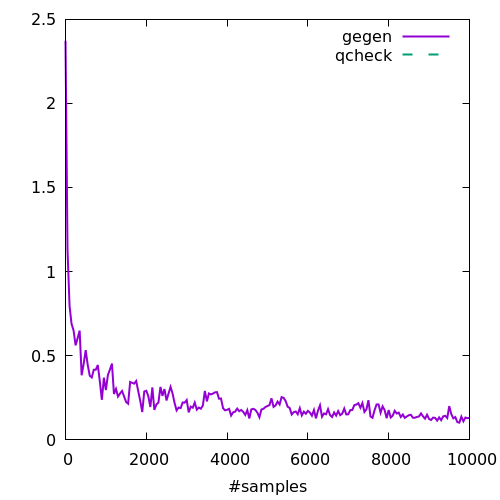}
    \caption{$n=20$\label{fig:evolf}}
  \end{subfigure}\\
\end{tabular}
\caption{Evolution of
  the generation time (in seconds) for sorted list of size $n$, with both QCheck
  and \textit{GeGen}\label{fig:evol}}
\end{figure}}

Figure~\ref{fig:evola} demonstrates that for small list sizes, QCheck
and \textit{GeGen} exhibit similar performance. In fact, QCheck
marginally outperforms \textit{GeGen}, as the rate of sorted lists
among lists of size 3 is relatively high. However, as $n$ increases,
the rejection rate also rises, causing \textit{GeGen} to surpass
QCheck. By the time $n=9$, \textit{GeGen} is already outperforming
QCheck by three orders of magnitude, as shown in
Figure \ref{fig:evolc}. Note that, due to the scale,
the \textit{GeGen} line appears very close to the x-axis.  For larger
values of $n$, rejection sampling via QCheck becomes impractical, and
thus only the results for \textit{GeGen} are presented. This figure
shows that the incremental nature of our generators allows them to
scale effectively, handling larger $n$ more efficiently.

%% file: related.tex
Random sampling of solutions of constrained systems is a well studied
subject and the literature is full of techniques for
SAT~\cite{Yu&al99, Pl&al19, Er&al12} models, and some results for CP
models exists, but mainly for finite domains
CSPs~\cite{De&al02,GoDe06,GoPe10,Va&al21,He&al22,Pe&al22}.

Closer to our work are~\cite{Cl&al14, La&al17, Zi&al21,
De&al19}. In \cite{Cl&al14}, the author present a technique for
automatically deriving test data generators from a predicate expressed
as a Boolean function. In order to speed up random generation they use
the lazy behaviour of the predicate to know its result on sets of
values, rather than individual values. Once they have computed a set
of values for which the predicate is going to return false, they
remove all of these values from the original set. They implement this
by relying on Haskell’s call-by-need semantics which applies the
predicate to a partially-defined value. This can be seen as
an \textit{ad-hoc} use of consistency on a partial assignment.
In \cite{Luck16}, the authors introduce Luck, a domain-specific
language for expressing property-based generators by annotating
logical predicates. These annotations allow users to control both the
distribution of generated values and the amount of constraint solving
performed before instantiating variables. Luck combines two
complementary generation strategies: incremental instantiation through
probabilistic narrowing and constraint solving to avoid making choices
that cannot lead to valid solutions. The authors provide a formal
semantics and prove soundness and completeness with respect to the
predicate semantics. While sharing the goal of integrating constraint
solving into property-based generation, our work differs in its
primary focus. Rather than providing a language to describe generators
and control the interplay between narrowing and constraint solving, we
focus on the construction of efficient representations of the solution
space for random sampling. In particular, our approach incrementally
refines these representations based on the observed sampling
behaviour, with the objective of achieving uniform sampling over the
solution space. Our contribution goes beyond integrating constraint
solving into random generation: we design a constraint-solving
procedure specifically tailored to constructing and incrementally
refining representations of the solution space, that is dedicated to
uniform random sampling.

In \cite{Zi&al21}, the authors propose a constraint-based generation
framework where constraint solving, which can be costly, is handled at
compilation time via preprocessing, while uniform sampling occurs at
runtime. While their approach is similar to ours, we find ours more
practical as it is incremental and does not require
preprocessing. Additionally, they do not focus on tuning the solving
method for uniform sampling, instead fixing the solver's depth and
size in advance, whereas we adapt constraint solving based on the
sampler’s output. More importantly, their approach cannot handle
recursive types, a limitation our method overcomes effectively.
In \cite{De&al19}, a Constraint Logic Programming approach for
Property based testing for Erlang programs, the authors generate
random tests for complex properties involving Modified
Condition/Decision Coverage, pattern matching, and higher-order
functions. This is equally useful but orthogonal to our wors as their
focus is put on code coverage and not of uniformity.  A similar idea,
in the field of computational statistics, is Adaptive Rejection
Sampling~\cite{MaMi10,Ma17} (ARS). The idea is to use a piecewise
linear density function instead of a single uniform envelope density
function.  Each time a sample has to be rejected, the rejected value
can be used to improve the piecewise approximation of the targeted
distribution. This therefore reduces the chance that the next attempt
will be rejected. This can however only be applied for sampling from
for specific families of densities.

Further from our field, other random sampling frameworks include the so-called
recursive method from Nijenhuis and Wilf~\cite{NW1978} and the framework of
Boltzmann sampling~\cite{DFLS2004}.
These frameworks provide a generic and efficient way to sample structured data
such as trees, algebraic data types, etc. but are not suitable for numerical data.
Finally, Monte-Carlo Markov Chains (MCMC) are another well-known tool for
sampling discrete structures~\cite{Ha70}.
However, here again, tuning a Markov Chain to make it efficient requires some
specific knowledge on the objects to sample, which make it unsuitable for
generically sampling into an arbitrary CSP solution space.

%% file: conclusion.tex
In this work, we have developped a method for building generators that
are unifom, efficient, and incremental.  We have understood that
neither the rejection sampling approach nor the constraint-based one
are universally optimal. Their relative efficiency depends on the
structure of the problem and, in particular, on the density of its
solutions. This motivates the need for a technique capable of
combining the simplicity of rejection sampling with the efficiency of
constraint-based generation, depending on the circumstances.  Our
method can be summarized as the gradual transition from rejection
sampling to a form of constraint-based one. Users have a generator
that adapts to their needs: If the generator is rarely used, there is
no need to spend time solving complex problems, and the underlying
search space remains largely unprocessed by the solver. As a result,
the generator behaves more like a rejection sampler. Conversely, if
the generator is heavily used, the solver will more aggressively
filter the search space, aiming to amortize the resolution cost
through improved generation speed. The techniques we propose holds for
both finite and continuous domains as the number of solutions in a
given search-space, whether it is finite or not, is abstracted by the
notion of weight. Also, our method for splitting the CSP into
independant pieces, which greatly improves our samplers, can be reused
in other context. For example, it makes the parallelization of the
solving process straightforward since each component can be treated
independently.

Our approach has several areas for improvement. We do not backtrack,
assuming that our exploration choices always enhance the sampler,
which is not guaranteed. Our generator synthesis is general and may be
less efficient than constraint-specific methods. Constrained data
structures often exhibit symmetries, suggesting that symmetry-breaking
techniques could be a natural extension. Additionally, non-uniform
methods may offer a better balance between bug-finding effectiveness
and computational cost. While uniformity can be expensive, surjective
generators may provide sufficient coverage. Finally, incorporating
corner-case analysis into solution sampling could further improve
results in future work.

%% file: biblio.bib
@article{Ta74,
title = {A note on finding the bridges of a graph},
journal = {Information Processing Letters},
volume = {2},
number = {6},
pages = {160-161},
year = {1974},
issn = {0020-0190},
doi = {https://doi.org/10.1016/0020-0190(74)90003-9},
url = {https://www.sciencedirect.com/science/article/pii/0020019074900039},
author = {R.Endre Tarjan}
}

@INPROCEEDINGS{Yu&al99,
  author={Yuan, J. and Shultz, K. and Pixley, C. and Miller, H. and Aziz, A.},
  booktitle={1999 IEEE/ACM International Conference on Computer-Aided Design. Digest of Technical Papers (Cat. No.99CH37051)}, 
  title={Modeling design constraints and biasing in simulation using BDDs}, 
  year={1999},
  pages={584-589},
  doi={10.1109/ICCAD.1999.810715}
}

@article{FiBi97,
author = {Fink, George and Bishop, Matt},
title = {Property-based testing: a new approach to testing for assurance},
year = {1997},
issue_date = {July 1997},
publisher = {Association for Computing Machinery},
address = {New York, NY, USA},
volume = {22},
number = {4},
issn = {0163-5948},
url = {https://doi.org/10.1145/263244.263267},
doi = {10.1145/263244.263267},
journal = {SIGSOFT Softw. Eng. Notes},
month = jul,
pages = {74–80},
numpages = {7}
}

@article{Cl&al00,
author = {Claessen, Koen and Hughes, John},
title = {QuickCheck: a lightweight tool for random testing of Haskell programs},
year = {2000},
issue_date = {Sept. 2000},
publisher = {Association for Computing Machinery},
address = {New York, NY, USA},
volume = {35},
number = {9},
issn = {0362-1340},
url = {https://doi.org/10.1145/357766.351266},
doi = {10.1145/357766.351266},
journal = {SIGPLAN Not.},
month = sep,
pages = {268–279},
numpages = {12}
}

@inproceedings{De&al02,
  title={Generating random solutions for constraint satisfaction problems},
  author={Rina Dechter and Kalev Kask and Eyal Bin and Roy Emek},
  booktitle={AAAI/IAAI},
  year={2002}
}

@InProceedings{Go&al04,
author="Gomes, Carla
and Sellmann, Meinolf",
editor="Wallace, Mark",
title="Streamlined Constraint Reasoning",
booktitle="Principles and Practice of Constraint Programming -- CP 2004",
year="2004",
publisher="Springer Berlin Heidelberg",
address="Berlin, Heidelberg",
pages="274--289",
isbn="978-3-540-30201-8"
}

@article{DFLS2004,
  title={Boltzmann samplers for the random generation of combinatorial structures},
  author={Duchon, Philippe and Flajolet, Philippe and Louchard, Guy and Schaeffer, Gilles},
  journal={Combinatorics, Probability \& Computing},
  volume={13},
  number={4-5},
  pages={577--625},
  year={2004},
  publisher={Cambridge University Press},
  doi={10.1017/S0963548304006315}
}

@InProceedings{Dy&al05,
author="Dybjer, Peter
and Haiyan, Qiao
and Takeyama, Makoto",
editor="Liu, Zhiming
and Araki, Keijiro",
title="Random Generators for Dependent Types",
booktitle="Theoretical Aspects of Computing - ICTAC 2004",
year="2005",
publisher="Springer Berlin Heidelberg",
address="Berlin, Heidelberg",
pages="341--355",
isbn="978-3-540-31862-0"
}

@inproceedings{GoDe06,
author = {Gogate, Vibhav and Dechter, Rina},
booktitle={International Conference on Principles and Practice of Constraint Programming},
year = {2006},
month = {09},
pages = {711-715},
title = {A New Algorithm for Sampling CSP Solutions Uniformly at Random},
isbn = {978-3-540-46267-5},
doi = {10.1007/11889205_56}
}

@article{Ge&al07,
title = {Random constraint satisfaction: Easy generation of hard (satisfiable) instances},
journal = {Artificial Intelligence},
volume = {171},
number = {8},
pages = {514-534},
year = {2007},
issn = {0004-3702},
doi = {https://doi.org/10.1016/j.artint.2007.04.001},
url = {https://www.sciencedirect.com/science/article/pii/S0004370207000653},
author = {Ke Xu and Frédéric Boussemart and Fred Hemery and Christophe Lecoutre}
}

@InProceedings{De&al08,
author="de Moura, Leonardo
and Bj{\o}rner, Nikolaj",
editor="Ramakrishnan, C. R.
and Rehof, Jakob",
title="Z3: An Efficient SMT Solver",
booktitle="Tools and Algorithms for the Construction and Analysis of Systems",
year="2008",
publisher="Springer Berlin Heidelberg",
address="Berlin, Heidelberg",
pages="337--340",
isbn="978-3-540-78800-3"
}

@article{MaMi10,
  title={A generalization of the adaptive rejection sampling algorithm},
  author={Luca Martino and Joaqu{\'i}n M{\'i}guez},
  journal={Statistics and Computing},
  year={2010},
  volume={21},
  pages={633-647},
  url={https://api.semanticscholar.org/CorpusID:1415606}
}

@article{GoPe10,
title = {A uniform random test data generator for path testing},
journal = {Journal of Systems and Software},
volume = {83},
number = {12},
pages = {2618-2626},
year = {2010},
note = {TAIC PART 2009 - Testing: Academic \& Industrial Conference - Practice And Research Techniques},
issn = {0164-1212},
doi = {https://doi.org/10.1016/j.jss.2010.08.021},
url = {https://www.sciencedirect.com/science/article/pii/S0164121210002244},
author = {Arnaud Gotlieb and Matthieu Petit}
}

@article{Er&al12,
  author       = {Stefano Ermon and
                  Carla P. Gomes and
                  Bart Selman},
  title        = {Uniform Solution Sampling Using a Constraint Solver As an Oracle},
  journal      = {CoRR},
  volume       = {abs/1210.4861},
  year         = {2012},
  url          = {http://arxiv.org/abs/1210.4861},
  eprinttype    = {arXiv},
  eprint       = {1210.4861},
  bibsource    = {dblp computer science bibliography, https://dblp.org}
}

@article{Ch&al13,
  author       = {Supratik Chakraborty and
                  Kuldeep S. Meel and
                  Moshe Y. Vardi},
  title        = {A Scalable and Nearly Uniform Generator of {SAT} Witnesses},
  journal      = {CoRR},
  volume       = {abs/1304.1584},
  year         = {2013},
  url          = {http://arxiv.org/abs/1304.1584},
  eprinttype    = {arXiv},
  eprint       = {1304.1584},
  bibsource    = {dblp computer science bibliography, https://dblp.org}
}

@InProceedings{Pe&al13,
author="Pelleau, Marie
and Min{\'e}, Antoine
and Truchet, Charlotte
and Benhamou, Fr{\'e}d{\'e}ric",
editor="Giacobazzi, Roberto
and Berdine, Josh
and Mastroeni, Isabella",
title="A Constraint Solver Based on Abstract Domains",
booktitle="Verification, Model Checking, and Abstract Interpretation",
year="2013",
publisher="Springer Berlin Heidelberg",
address="Berlin, Heidelberg",
pages="434--454",
isbn="978-3-642-35873-9"
}

@InProceedings{Cl&al14,
author="Claessen, Koen
and Dureg{\aa}rd, Jonas
and Pa{\l}ka, Micha{\l} H.",
editor="Codish, Michael
and Sumii, Eijiro",
title="Generating Constrained Random Data with Uniform Distribution",
booktitle="Functional and Logic Programming",
year="2014",
publisher="Springer International Publishing",
address="Cham",
pages="18--34",
isbn="978-3-319-07151-0"
}

@article{La&al17,
author = {Lampropoulos, Leonidas and Paraskevopoulou, Zoe and Pierce, Benjamin C.},
title = {Generating Good Generators for Inductive Relations},
year = {2017},
issue_date = {January 2018},
publisher = {Association for Computing Machinery},
address = {New York, NY, USA},
volume = {2},
number = {POPL},
url = {https://doi.org/10.1145/3158133},
doi = {10.1145/3158133},
journal = {Proc. ACM Program. Lang.},
month = {dec},
articleno = {45},
numpages = {30}
}

@article{Ma17,
author = {Martino, Luca},
year = {2017},
month = {06},
pages = {},
title = {Parsimonious Adaptive Rejection Sampling},
volume = {53},
journal = {Electronics Letters},
doi = {10.1049/el.2017.1711}
}

@inproceedings{Zi&al18,
  TITLE = {{Finding solutions by finding inconsistencies}},
  AUTHOR = {Ziat, Ghiles and Pelleau, Marie and Truchet, Charlotte and Min{\'e}, Antoine},
  URL = {https://hal.science/hal-01885769},
  BOOKTITLE = {{CP 2018 - 24th International Conference on Principles and Practice of Constraint Programming}},
  ADDRESS = {Lille, France},
  PAGES = {1-16},
  YEAR = {2018},
  MONTH = Aug,
  HAL_ID = {hal-01885769},
  HAL_VERSION = {v1},
}

@INPROCEEDINGS{Pl&al19,
  author={Plazar, Quentin and Acher, Mathieu and Perrouin, Gilles and Devroey, Xavier and Cordy, Maxime},
  booktitle={2019 12th IEEE Conference on Software Testing, Validation and Verification (ICST)}, 
  title={Uniform Sampling of SAT Solutions for Configurable Systems: Are We There Yet?}, 
  year={2019},
  pages={240-251},
  doi={10.1109/ICST.2019.00032}
}

@InProceedings{De&al19,
author="De Angelis, Emanuele
and Fioravanti, Fabio
and Palacios, Adri{\'a}n
and Pettorossi, Alberto
and Proietti, Maurizio",
editor="Beyer, Dirk
and Keller, Chantal",
title="Property-Based Test Case Generators for Free",
booktitle="Tests and Proofs",
year="2019",
publisher="Springer International Publishing",
address="Cham",
pages="186--206",
isbn="978-3-030-31157-5"
}

@InProceedings{Va&al21,
  author =	{Vavrille, Mathieu and Truchet, Charlotte and Prud'homme, Charles},
  title =	{{Solution Sampling with Random Table Constraints}},
  booktitle =	{27th International Conference on Principles and Practice of Constraint Programming (CP 2021)},
  pages =	{56:1--56:17},
  series =	{Leibniz International Proceedings in Informatics (LIPIcs)},
  ISBN =	{978-3-95977-211-2},
  ISSN =	{1868-8969},
  year =	{2021},
  volume =	{210},
  editor =	{Michel, Laurent D.},
  publisher =	{Schloss Dagstuhl -- Leibniz-Zentrum f{\"u}r Informatik},
  address =	{Dagstuhl, Germany},
  URL =		{https://drops.dagstuhl.de/opus/volltexte/2021/15347},
  URN =		{urn:nbn:de:0030-drops-153477},
  doi =		{10.4230/LIPIcs.CP.2021.56}
}

@InProceedings{Zi&al21,
  author =	{Ziat, Ghiles and Dien, Matthieu and Botbol, Vincent},
  title =	{{Automated Random Testing of Numerical Constrained Types}},
  booktitle =	{27th International Conference on Principles and Practice of Constraint Programming (CP 2021)},
  pages =	{59:1--59:19},
  series =	{Leibniz International Proceedings in Informatics (LIPIcs)},
  ISBN =	{978-3-95977-211-2},
  ISSN =	{1868-8969},
  year =	{2021},
  volume =	{210},
  editor =	{Michel, Laurent D.},
  publisher =	{Schloss Dagstuhl -- Leibniz-Zentrum f{\"u}r Informatik},
  address =	{Dagstuhl, Germany},
  URL =		{https://drops.dagstuhl.de/opus/volltexte/2021/15350},
  URN =		{urn:nbn:de:0030-drops-153502},
  doi =		{10.4230/LIPIcs.CP.2021.59}
}

@article{St&al21,
author = {Steele Jr., Guy L. and Vigna, Sebastiano},
title = {LXM: Better Splittable Pseudorandom Number Generators (and Almost as Fast)},
year = {2021},
issue_date = {October 2021},
publisher = {Association for Computing Machinery},
address = {New York, NY, USA},
volume = {5},
number = {OOPSLA},
url = {https://doi.org/10.1145/3485525},
doi = {10.1145/3485525},
journal = {Proc. ACM Program. Lang.},
month = {oct},
articleno = {148},
numpages = {31}
}

@InProceedings{Pe&al22,
author="Pesant, Gilles
and Quimper, Claude-Guy
and Verhaeghe, H{\'e}l{\`e}ne",
editor="Schaus, Pierre",
title="Practically Uniform Solution Sampling in Constraint Programming",
booktitle="Integration of Constraint Programming, Artificial Intelligence, and Operations Research",
year="2022",
publisher="Springer International Publishing",
address="Cham",
pages="335--344",
isbn="978-3-031-08011-1"
}

@INPROCEEDINGS{He&al22,
  author={He, Kun and Wang, Chunyang and Yin, Yitong},
  booktitle={2022 IEEE 63rd Annual Symposium on Foundations of Computer Science (FOCS)}, 
  title={Sampling Lovász local lemma for general constraint satisfaction solutions in near-linear time}, 
  year={2022},
  pages={147-158},
  doi={10.1109/FOCS54457.2022.00021}
}

@article{Huffman1952,
  title={A method for the construction of minimum-redundancy codes},
  author={Huffman, David Albert},
  journal={Proceedings of the IRE},
  volume={40},
  number={9},
  pages={1098--1101},
  year={1952},
  publisher={IEEE}
}

@book{NW1978,
  author={Nijenhuis, Albert and Wilf, Herbert},
  title={Combinatorial Algorithms: For Computers and Hard Calculators},
  year={1978},
  isbn={0125192606},
  publisher={Academic Press, Inc.},
  address={USA},
  edition={2nd},
  doi={10.1016/C2013-0-11243-3},
}

@article{Ha70,
 ISSN = {00063444, 14643510},
 URL = {http://www.jstor.org/stable/2334940},
 author = {W. K. Hastings},
 journal = {Biometrika},
 number = {1},
 pages = {97--109},
 publisher = {[Oxford University Press, Biometrika Trust]},
 title = {Monte Carlo Sampling Methods Using Markov Chains and Their Applications},
 urldate = {2025-03-31},
 volume = {57},
 year = {1970}
}

@misc{qcheck,
  author       = {Simon Cruanes},
  title        = {QCheck: Property-based testing for OCaml},
  howpublished = {\url{https://github.com/c-cube/qcheck}},
  note         = {Accessed: 2025-04-02}
}

@article{Luck16,
  author       = {Leonidas Lampropoulos and
                  Diane Gallois{-}Wong and
                  Catalin Hritcu and
                  John Hughes and
                  Benjamin C. Pierce and
                  Li{-}yao Xia},
  title        = {Beginner's Luck: {A} Language for Property-Based Generators},
  journal      = {CoRR},
  volume       = {abs/1607.05443},
  year         = {2016},
  url          = {https://doi.org/10.48550/arXiv.1607.05443},
  doi          = {10.48550/ARXIV.1607.05443},
  eprinttype   = {arXiv},
  eprint       = {1607.05443},
  bibsource    = {dblp computer science bibliography, https://dblp.org}
}
